\documentclass{andp2012}
\usepackage[english]{babel}
\usepackage{units, bm}
\usepackage{hyperref}
\usepackage{cleveref}

\category{Feature Article}

\title{Carrier dynamics in graphene: ultrafast many-particle phenomena}

\author[E.\, Malic]{Ermin Malic\inst{1,}\footnote{Corresponding author\quad E-mail:~\textsf{ermin.malic@chalmers.se}}}
\author[T.\, Winzer]{Torben Winzer\inst{2}}
\author[F.\, Wendler]{Florian Wendler\inst{2}}
\author[S.\, Brem]{Samuel Brem\inst{1}}
\author[R.\, Jago]{Roland Jago\inst{1}}
\author[A.\, Knorr]{Andreas Knorr\inst{2}$\phantom{asdfasdfasdfasdfasdfasdfasdfasdfasdfasdfasdfasdfasdfasdfa}$}

\author[M.\, Mittendorff]{Martin Mittendorff\inst{3}}
\author[J.\, K\"{o}nig-Otto]{Jacob C. K\"{o}nig-Otto\inst{3}}
\author[T.\, Pl\"{o}tzing]{Tobias Pl\"{o}tzing\inst{4}}
\author[D.\, Neumaier]{Daniel Neumaier\inst{4}}
\author[H. \, Schneider]{Harald Schneider\inst{3}}
\author[M.\, Helm]{Manfred Helm\inst{3}}
\author[S.\, Winnerl]{Stephan Winnerl\inst{3}}
\address[1]{Department of Physics, Chalmers University of Technology, 41296 G\"{o}tborg, Sweden}
\address[2]{Institut f\"{u}r Theoretische Physik, Technische Universit\"{a}t Berlin, 10623 Berlin, Germany}
\address[3]{Helmholtz-Zentrum Dresden-Rossendorf, 01314 Dresden, Germany}
\address[4]{Advanced Microelectronic Center Aachen, AMO GmbH, 52074 Aachen, Germany}
\shortauthors{E. Malic et al.}

\begin{abstract}
Graphene  is an ideal material to study fundamental Coulomb- and phonon-induced carrier scattering processes. Its remarkable gapless and linear band structure opens up new carrier relaxation channels. In particular, Auger scattering  bridging the valence and the conduction band changes the number of charge carriers and gives rise to a significant carrier multiplication - an ultrafast many-particle phenomenon that is promising for the design of highly efficient photodetectors. Furthermore, the vanishing density of states at the Dirac point combined with ultrafast phonon-induced intraband scattering results in an accumulation of carriers and a population inversion suggesting the design of graphene-based terahertz lasers. 
Here, we review our work on the ultrafast carrier dynamics in graphene and Landau-quantized graphene is presented providing  a microscopic view on the appearance of carrier multiplication and population inversion. 
\end{abstract}
\shortabstract

\begin{document}
\maketitle

\section{Introduction}

\begin{figure}[t!]
  \begin{center}
     \includegraphics[width=\columnwidth]{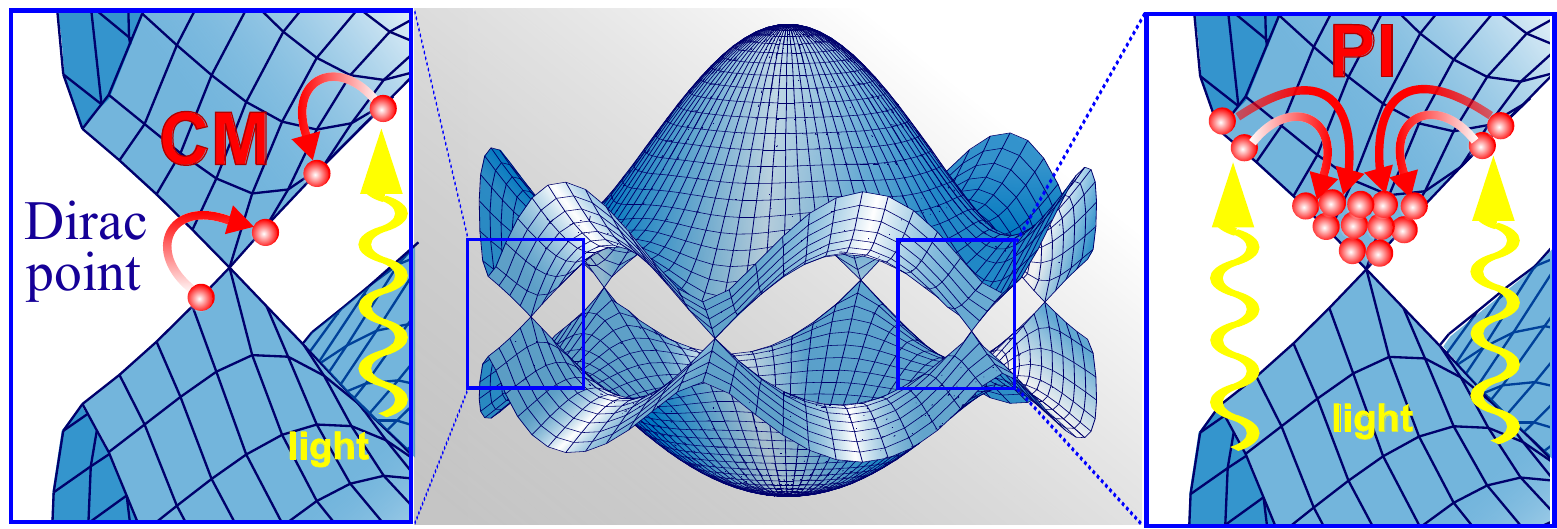}
  \end{center}
  \caption{\textbf{Dirac cones and ultrafast phenomena.} (a) Electronic band structure of graphene is shown over the entire Brillouin zone. At the edges,  Dirac points appear that are characterized by linear and gapless bands giving rise to 
 technologically promising carrier multiplication (CM) and population inversion (PI), cf. the zoom-ins into the region around a Dirac point.  Figure adapted from Ref. \cite{malic13}.
}\label{fig1_bands} 
\end{figure}

The remarkable electronic properties of graphene \cite{geim07, castro_neto09} give rise to an ultrafast carrier dynamics that is highly interesting both from the perspective of fundamental research as well as technological applications \cite{bonaccorso10, avouris12, malic13, malic16, malic16b}.
The electronic band structure is characterized by Dirac cones appearing at the edges of the Brillouin zone and showing linear and gapless bands (Fig. \ref{fig1_bands}) - in contrast to conventional parabolic and gapped semiconductors. This has a strong impact on possible carrier relaxation processes and in particular opens up new relaxation channels. A prominent example are Auger processes, where one electron bridges the valence and the conduction band, while the other involved electron remains in the same band, cf. Fig.  \ref{fig1_bands}. These processes
changes the number of charge carriers and result in a technologically promising many-particle phenomenon called carrier multiplication (CM) \cite{winzer10,winzer12b, brida13, polini13, ploetzing14, gierz15, koppens13, kadi15, hofmann15,  wendler14}.  Another remarkable consequence of graphene's linear band structure and the vanishing density of states at the Dirac point is the possibility of accumulating carriers at low energies resulting in a population inversion (PI), i.e. carrier occupations higher than 0.5 in the conduction band \cite{li12, winzer13, gierz13, jago15, wendler15b, brem16}. The appearance of PI in graphene demonstrates its applicability as a gain medium for lasers that could also operate in the technologically promising terahertz spectral range.

Carrier multiplication holds the potential to increase the power conversion efficiency of photodetecting and photovoltaic
devices \cite{koppens14}. However, due to the absence of a band gap the extraction of charge carriers remains a substantial challenge.
A possible strategy to circumvent this drawback is Landau quantization of graphene: the presence of magnetic fields induces a collapse of the Dirac cone into discrete non-equidistant Landau levels (LLs) that can be selectively addressed by circularly polarized light \cite{goerbig11, wendler15}, cf. Fig. \ref{fig1_bands}. The externally tunable gaps between the LLs also suggest the possibility of tunable terahertz Landau-level lasers.

In this Feature article, we present a review of our  joint theory-experimental work on the carrier dynamics in graphene and Landau-quantized graphene is presented. In particular, we provide a microscopic view on the elementary many-particle processes behind the generation and decay of carrier multiplication and population inversion in optically excited graphene in and without the presence of magnetic fields. Note that a separate feature article with a focus on the experimental investigation of the carrier dynamics at very low energies is published in this special issue [S. Winnerl et al., citation when available].

\section{Theoretical approach}
The carrier dynamics in graphene has been extensively investigated in high-resolution pump-probe experiments  \cite{dawlaty08,george08, sun08,   plochocka09, choi09, kumar09,newson09, shang10, wang10_dts, huang10, breusing11, winnerl11,  strait11,brida13, ploetzing14}  measuring the differential transmission (DTS) as well as time- and angle-resolved
photoemission (ARPES) experiments  \cite{gierz13, hofmann13, gierz15, hofmann15, gierz17}. To provide microscopic insights into elementary scattering processes behind the carrier dynamics and the observed ultrafast phenomena, we have developed a microscopic many-particle approach allowing us to temporally and spectrally track the way of optically excited electrons in graphene \cite{malic13, malic11b}. 

The core of our approach is built by Graphene Bloch Equations (GBE) that have been derived in the density matrix formalism within the second-order Born-Markov approximation \cite{haug04, rossi02, knorr96, kira06}. We obtain a system of coupled equations of motion for the carrier occupation $\rho^{\lambda}_{\bf k}=\langle a_{\lambda\bf k}^{+}a^{\phantom{+}}_{\lambda\bf k}\rangle$ in the state $(\bf{k}, \lambda)$ characterized by the momentum $\bf k$ and the band index $\lambda$, the microscopic polarization $p_{\bf k}=\langle a_{v\bf k}^{+}a^{\phantom{+}}_{c\bf k}\rangle $ that is a measure for the transition probability between the valence ($\lambda=v$) and the conduction ($\lambda=c$) band, and phonon number $n_{\bf q}^j=\langle b_{j\bf q}^{+}b^{\phantom{+}}_{j\bf q}\rangle$ in the considered optical or acoustic phonon mode $j$ with the phonon momentum $\bf q$. 
Here, we have expressed the microscopic quantities in the formalism of second quantization introducing creation and annihilation operators for electrons ($a_{{\lambda\bf k}}^{+}, a_{{\lambda\bf k}}^{\phantom{+}}$) and phonons ($b_{{j\bf q}}^{+}, b_{{j\bf q}}^{\phantom{+}}$). 

\begin{figure}[t!]
  \begin{center}
     \includegraphics[width=\columnwidth]{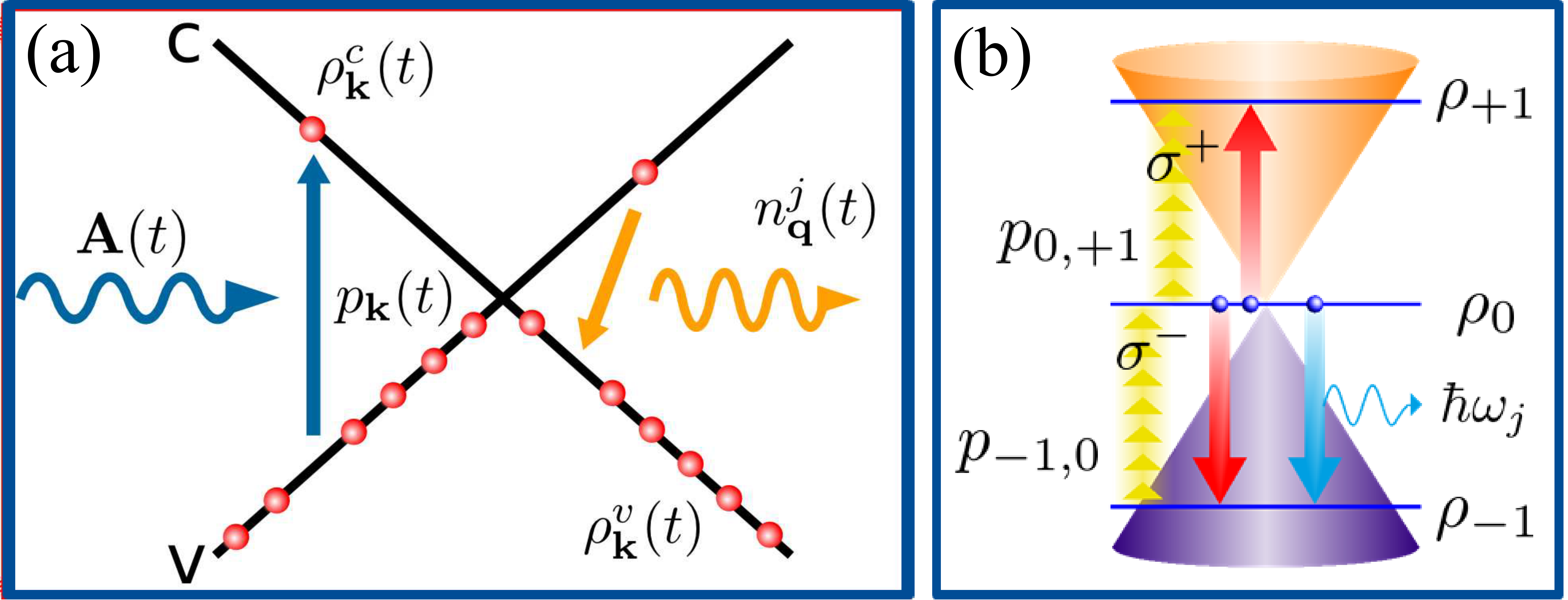}
  \end{center}
  \caption{\textbf{Relaxation channels.} Illustration of microscopic quantities and relaxation channels determining the carrier dynamics of optically excited carriers in (a) graphene and (b) Landau-quantized graphene. Red (blue) arrows in (b) denote Auger and carrier-phonon scattering, respectively. Figure a adapted from Ref. \cite{malic11b}.
}\label{fig2_dmt} 
\end{figure}

Figure \ref{fig2_dmt}(a) illustrates the introduced microscopic quantities in a non-equilibrium situation: First, graphene is excited by an optical pulse that is characterized by the vector potential $\bf A(t)$. The process of optical excitation is described by the  microscopic polarization $p_{\bf k}$. The excitation and the subsequent scattering dynamics changes the occupation probabilities $\rho^c_{\bf k}$ and $\rho^v_{\bf k}$ in the involved bands. The carrier scattering with phonons transfers the optically injected energy to the lattice system and changes the phonon number $n^j_{\bf q}$. 
Therefore, solving the Graphene Bloch Equations, we can track the way of carriers in time and momentum from the optical excitation via carrier-carrier and carrier-phonon scattering towards an equilibrium distribution.

Applying the Heisenberg equation of motion and exploiting the fundamental commutation relations for fermions and bosons, we derive the GBE yielding \cite{malic13}
\begin{eqnarray}
\label{bloch_p}
 \dot{p}_{\bf k}&=&i\Delta\omega_{\bf k}p_{\bf k}-i\Omega_{\bf k}^{vc}\big(\rho_{\bf k}^c-\rho_{\bf k}^v\big)+\mathcal{U}_{\bf k}-\gamma_{\bf k} p_{\bf k},\\[8pt]
\label{bloch_rho}
 \dot{\rho}_{\bf k}^{\lambda}&=&\pm2{\rm{Im}}\big(\Omega_{\bf k}^{vc*}p_{\bf k}\big)+\Gamma_{\lambda,{\bf k}}^{\rm{in}}\big(1+\rho_{\bf k}^\lambda\big)-\Gamma_{\lambda,{\bf k}}^{\rm{out}}\rho_{\bf k}^\lambda,\\[8pt]
\label{bloch_nph}
\dot{n}_{\bf q}^j&=&\Gamma_{j,{\bf q}}^{\rm{em}}\big(1+n_{\bf q}^j\big)-\Gamma_{j,{\bf q}}^{\rm{abs}}n_{\bf q}^j-\gamma_j\big(n_{\bf q}^j-n_B\big)
\end{eqnarray}
with the energy difference $\hbar\Delta\omega_{\bf k}=(\varepsilon_{\bf
  k}^{v}-\varepsilon_{\bf k}^{c})$,  the Bose-Einstein distribution $n_B$ denoting the equilibrium distribution of phonons,  the phonon lifetime $\gamma^{-1}_j$, and the Rabi frequency
$\Omega^{vc}_{\bf k}(t)=i\frac{e_{0}}{m_{0}} {\bf{M}}_{{\bf k}}^{vc}\cdot {\bf{A}}(t)$ describing the optical excitation of graphene with ${\bf{M}}_{{\bf k}}^{vc}$ as the optical matrix element. 
The intraband Rabi frequency $\Omega^{\lambda\lambda}_{\bm k}(t)$ contributes to a renormalization of the band structure, but since this effect is very small it has been neglected here.  The appearing matrix elements are calculated with full tight-binding wave functions in the nearest-neighbor approximation \cite{malic13}, which is known to be a good approximation for graphene close to the Dirac point \cite{reich02b}.  All quantities in GBE (except for $n_B$ and $\Delta\omega_{\bf k}$) depend on time.
 The equations take into account all relevant two-particle relaxation channels including Coulomb- and phonon-induced intra- and interband as well as intra- and intervalley scattering processes.
The time- and momentum-dependent scattering rates $\Gamma_{\lambda{\bf k}}(t)=\Gamma_{\lambda{\bf k}}^{\rm{cc}}(t)+\Gamma_{\lambda{\bf k}}^{\rm{cp}}(t)$  entering the equation for the carrier occupation $\rho_{\bf k}^\lambda (t)$   describe the strength of the  carrier-carrier (cc) and carrier-phonon (cp) scattering processes. For the phonon occupation, we obtain the corresponding emission and absorption rates $\Gamma_{j,{\bf q}}^{\rm{em/abs}}$ \cite{malic11b}. The many-particle scattering also contributes to  diagonal ($\gamma_{\bf k}(t)$) and off-diagonal dephasing ($\mathcal{U}_{\bf k}(t)$) of the microscopic polarization $p_{\bf k}(t)$. More details on the theoretical approach can be found in the appendix.

The magnetic field is incorporated into the equations by exploiting the Peierls substitution accounting for the change of electron momentum induced by the confinement into cyclotron orbits \cite{goerbig11,sipe12}. 
This results in a drastic change of the electronic band structure, where the Dirac cone collapses into discrete non-equidistant
Landau levels \cite{goerbig11,sadowski06} with 
$
 \varepsilon_n = \rm{sgn}\{n\}\hbar v_F \sqrt{\frac{2 e_0 B}{\hbar} \lvert n \rvert}.
$
 Here, the magnetic field $B$ is perpendicular to the graphene layer, $v_F$ denotes the Fermi velocity 
in graphene, and $n=...,-2,-1,0,1,2, ...$ is the LL quantum number. Optical selection rules
only allow transitions with quantum
numbers \cite{sipe12, wendler15} $\lvert n\rvert \rightarrow \lvert n  \rvert \pm 1$ induced by left (-) or right (-) circularly polarized light that is denoted by $\sigma_{\pm}$.  Figure \ref{fig2_dmt}(b) shows the three energetically lowest LLs $n=-1,0,+1$, the allowed optical excitation with $\sigma_{\pm}$ light as well as the possible Auger scattering processes between the energetically equidistant LLs (blue arrows) and carrier-phonon scattering. The latter results in emission or absorption of a phonon with the energy $\hbar \omega_j$ that is resonant to the LL separation. In analogy to the case without the magnetic field, we define the microscopic polarization $p_{n,n'}$ and carrier occupation $\rho_n $ that now only depend on  LL indices $n, n'$. More details on the Bloch equations in Landau-quantized graphene can be found in Ref. \cite{wendler15}.

\section{Carrier dynamics}

\begin{figure*}[t!]
  \begin{center}
     \includegraphics[width=0.65\linewidth]{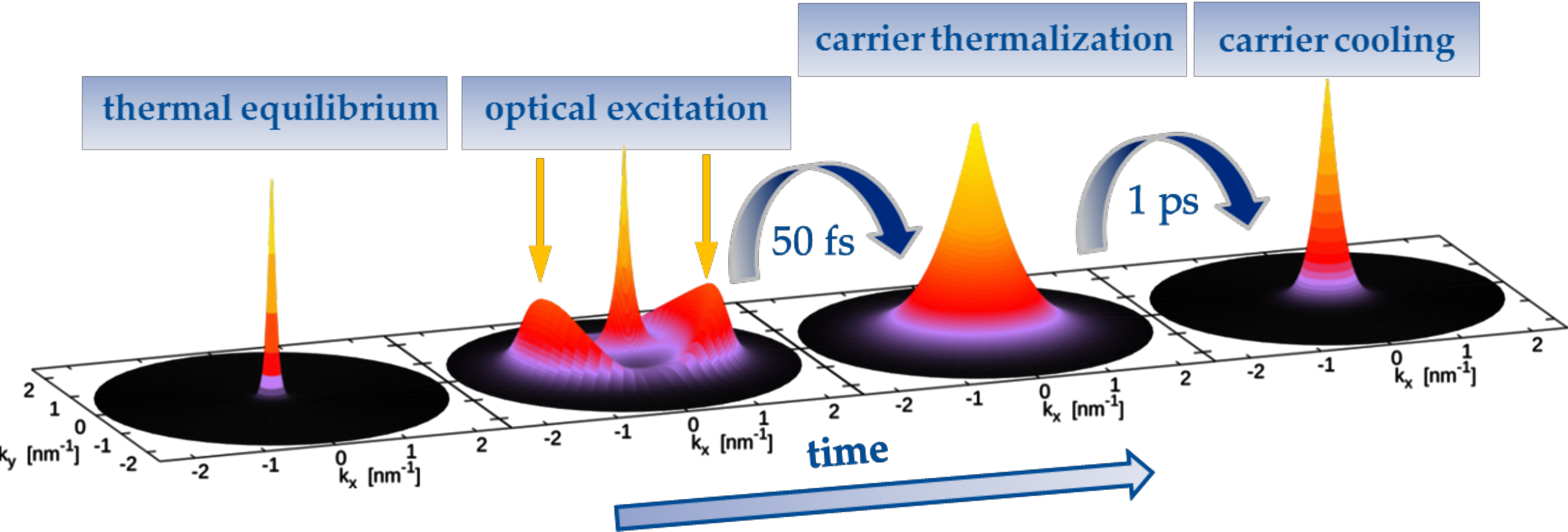}
  \end{center}
  \caption{\textbf{Carrier dynamics in graphene.} Illustration of the main steps during the carrier relaxation dynamics in graphene. Carrier occupation $\rho_{\bf k}^c(t)$ in the conduction band is plotted as a function of the momentum ${\bf k}=(k_x, k_y)$ for different exemplary times. Figure adapted from Ref. \cite{winzer13b}.
}\label{fig3_dynamics} 
\end{figure*}

By solving the Graphene Bloch Equations, we can track the way of optically excited carriers as a function of time and momentum along the Dirac cone or Landau levels. We can resolve fundamental many-particle relaxation channels including Coulomb- and phonon-induced intra- and interband processes driving the carriers towards an equilibrium Fermi distribution.
Figure \ref{fig3_dynamics} illustrates the main steps characterizing the carrier dynamics in graphene:  thermal distribution,  optical excitation, carrier thermalization, and  carrier cooling. (i) The starting point is a thermal Fermi distribution that is  determined by the initial temperature. (ii) Next, a non-equilibrium carrier distribution is generated by applying an optical pulse described by a Gaussian-shaped vector potential ${\bf A}(t)=A_0\,\hat{\bf e}_{\mathcal{P}}\,\exp\left(-t^2/(2\sigma_t^2)\right)\cos(\omega_L t)$,
where $A_0$ is the amplitude determining the pump fluence, $\hat{\bf e}_{\mathcal{P}}=(\cos\phi_{\mathcal{P}},\sin\phi_{\mathcal{P}})$ the polarization unit vector, $\sigma_t$ the pulse duration, and $\hbar \omega_L$ the excitation energy. Here, we apply an energy of $\hbar\omega_L=\unit[1.5]{eV}$, a pulse duration of $\sigma_t=\unit[10]{fs}$, and a pump fluence of $\unit[1]{\rm{\mu} J/cm^2}$. The optical excitation generates a highly anisotropic non-equilibrium distribution that is centered around $k\approx \pm \unit[1.25]{nm^{-1}}$ corresponding to the excitation energy of $\unit[1.5]{eV}$. The anisotropy can be ascribed to the anisotropic optical matrix element ${\bf M}^{vc}_{\bf k}$ \cite{malic12, winnerl14}. The importance of anisotropy for the carrier dynamics in graphene is discussed in detail in a separate feature article in this special issue [S. Winnerl et al., citation when available]. 

(iii) The non-equilibrium electrons are redistributed to energetically lower states via combined carrier-carrier and carrier-phonon scattering resulting in a thermalized carrier distribution already after approximately $\unit[50]{fs}$. Here, carriers are in equilibrium among each other, however their energy is much higher compared to the initial state, i.e. the reached hot thermalized Fermi distribution is spectrally much broader than the initial thermal distribution. We find carrier temperatures above \unit[1000]{K} strongly depending on the applied pump fluence. In this step, the initially highly anisotropic carrier distribution also becomes isotropic again. This can be ascribed to highly efficient scattering with optical phonons \cite{malic11b, breusing11, winnerl11}. 

(iv) Finally,  phonon-induced scattering redistributes energy from the excited carrier system to the lattice on a picosecond timescale resulting in a narrowing of the carrier distribution. Eventually, the initial thermal Fermi distribution is reached. 
The process of thermalization and carrier cooling cannot be strictly separated in time, since carrier-phonon scattering contributes to both the thermalization and the energy dissipation. Nevertheless, high-resolution pump-probe experiments measuring the differential transmission signal have found two distinct decay times with $\tau_1$ typically in the range of $\unit[100]{fs}$ and $\tau_2$ typically around $\unit[1]{ps}$  representing thermalization and cooling times \cite{dawlaty08,george08, sun08,   plochocka09, choi09, kumar09,newson09, shang10, wang10_dts, huang10, breusing11, winnerl11,  strait11,brida13, hofmann13}.

In Landau-quantized graphene, the non-equidistant separation of Landau levels is expected to strongly suppress elastic Coulomb scattering processes. The latter are only allowed, if two Landau level transitions with the same energy separation are available facilitating energy conservation. In the next section, we will show that in spite of this restriction Auger scattering processes play a crucial role for the carrier dynamics in Landau-quantized graphene, since one specific finds Landau levels with the same energetic spacing.  Phonon-induced
transitions between Landau levels are strongly suppressed unless the respective transition is in resonance
with the optical phonon energy, at least  up to a detuning determined by the impurity-induced Landau level broadening. 
Since the LL distance can be tuned by the magnetic field, the efficiency of scattering with optical phonons strongly  depends
on the magnetic field strength \cite{wendler13}. Since the velocity of acoustic phonons is much smaller than the electronic Fermi velocity, the impact of acoustic phonons is generally very small within the energetically lowest Landau levels, as long as the LL broadening is not larger than \unit[10]{meV} \cite{wendler15}. 

Carrier thermalization and carrier cooling discussed above are typical relaxation steps during the carrier dynamics in any material. There are differences in the time scales, but the qualitative behavior is similar. However, graphene exhibits some specific technologically promising ultrafast phenomena characterizing its dynamics. This includes the appearance of a significant carrier multiplication and transient population inversion. These many-particle phenomena will be discussed in detail in the case of graphene and Landau-quantized graphene. Note that the section on the carrier multiplication is based on Ref. \cite{malic16b}.

\section{Carrier multiplication}

Carrier multiplication is a many-particle phenomenon describing the generation of multiple electron-hole pairs  internal scattering \cite{nozik02}. The underlying physical mechanism is Auger scattering, which is a specific collinear Coulomb-induced interband relaxation channel, cf. Figs. \ref{fig1_bands} and \ref{fig4_cm}. In contrast to all other Coulomb processes,  Auger scattering changes the charge carrier density consisting of electrons in the conduction and holes in the valence band. We distinguish Auger recombination (AR) and the inverse process of impact excitation (IE) \cite{winzer10, winzer12b}. While IE increases the charge carrier density, the AR reduces the number of carriers through recombination of excited electrons with holes in the valence band. The efficiency of the two Auger processes depends on the excitation regime and the resulting Pauli blocking. After a weak optical excitation, the probability of the interband process close to the Dirac point is much higher for IE, since here an electron from the almost full valence band scatters into the weakly populated conduction band. The inverse process is strongly suppressed by Pauli blocking resulting in an overall  multiplication of optically excited carriers \cite{winzer10,winzer12b}. This carrier multiplication has the potential
to increase the responsivity in photodetecting devices as well as power conversion efficiency in photovoltaic devices \cite{koppens14}.

While carrier multiplication is possible in any two-dimensional material with a sufficiently small band gap, the linear band structure of graphene is favorable, since it facilitates the conservation of energy and momentum at the same time. 
 In graphene, the occurrence of Auger-induced carrier multiplication
has been theoretically predicted \cite{winzer10,winzer12b,polini13,basko13, winzer16}
and experimentally confirmed \cite{brida13,ploetzing14,gierz15}. 
Furthermore, in spite of the non-equidistant separation of Landau levels, Auger scattering has been theoretically and experimentally demonstrated as the most important relaxation channel even in Landau-quantized graphene \cite{wendler14, wendler15b, winnerl15} resulting in a significant carrier multiplication \cite{wendler14}. 

\subsection{Carrier multiplication in graphene}

\begin{figure}[t!]%
\centering
\includegraphics*[width=0.75\columnwidth]{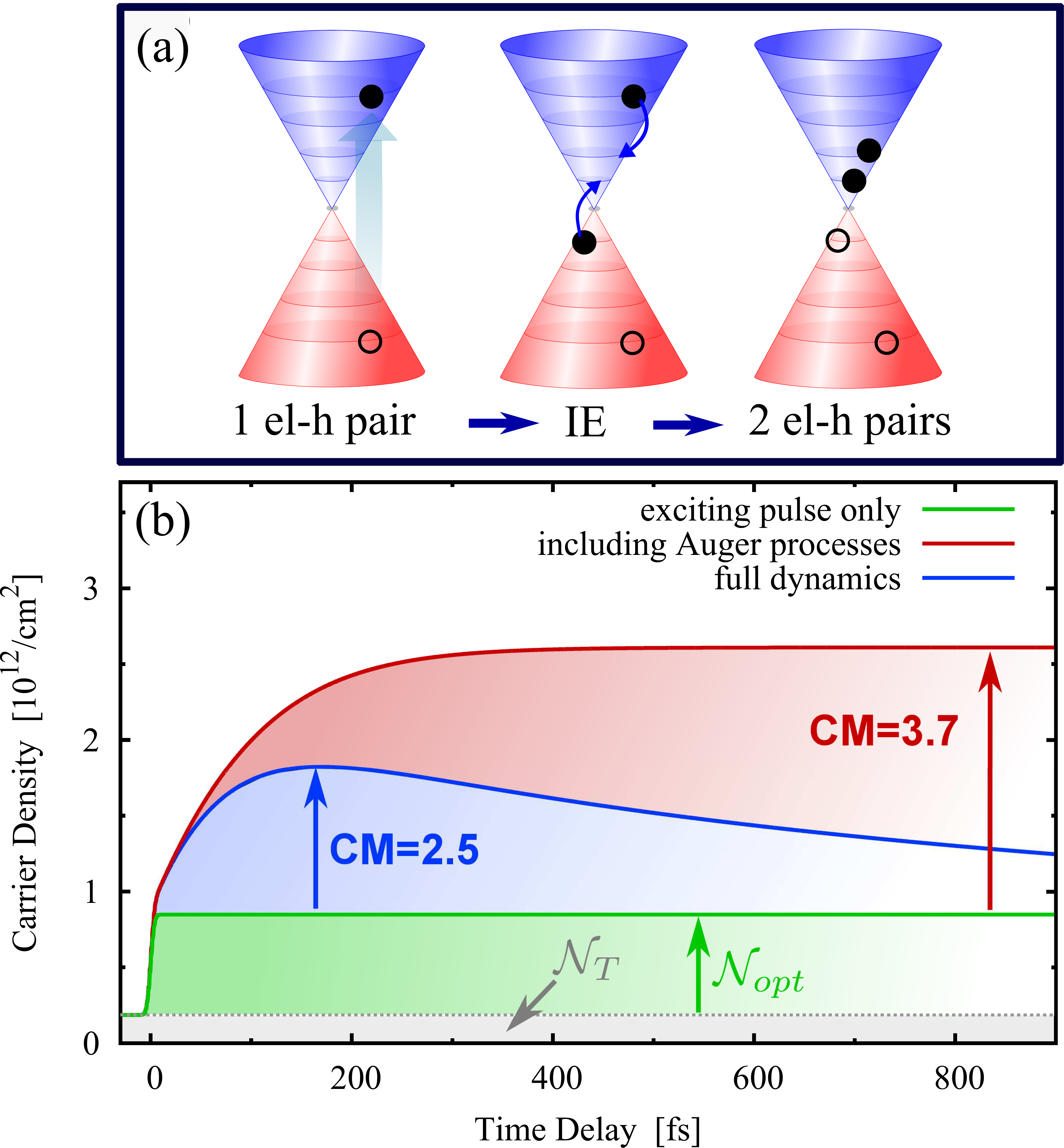}
\caption{\textbf{Carrier multiplication in graphene.} 
(a) Schematic illustration of the impact excitation (IE) multiplying the optically excited electron-hole pairs through internal scattering.
(b) Temporal evolution of the carrier density ${\mathcal{N}}(t)$ taking into account only carrier-light interaction (green line), only  Coulomb scattering including Auger processes (red), and full dynamics including carrier-phonon  channels (blue line).
 Auger scattering leads to a multiplication of the initially excited carrier density $\mathcal{N}_{\rm{opt}}$ by a factor of $3.7$. Phonon-induced recombination counteracts the effect and reduces the CM factor to $2.5$. Figure b taken from Ref. \cite{torben_phd}}.
\label{fig4_cm}
\end{figure}
Auger scattering-induced multiplication of electron-hole pairs after absorption of a single photon can be quantified by a time-dependent carrier multiplication factor 
\begin{equation}
 \rm{CM}(t)=\frac{\mathcal{N}(t)-\mathcal{N}_{T}}{\mathcal{N}_{\rm{opt}}(t)-\mathcal{N}_{T}},\label{eq_CMdef}
\end{equation}
where ${\mathcal{N}}(t)$, ${\mathcal{N}}_{\rm{opt}}(t)$ and ${\mathcal{N}}_{T}$ are the total, the purely optically excited, and the thermal carrier density, respectively.

To investigate whether a carrier multiplication appears in graphene, we calculate the temporal evolution of the carrier density ${\mathcal{N}}(t)=\frac{1}{A}\sum_{\bf k, \lambda}\rho^\lambda_{\bf k}$, cf. Fig. \ref{fig4_cm}. To understand the underlying microscopic mechanism, we subsequently switch on different scattering channels. First, we apply an optical pulse characterized by a pump fluence of $\unit[2.3]{\rm{\mu} J cm^{-2}}$, a pulse width of $\unit[10]{fs}$, and an excitation energy of $\unit[1.5]{eV}$.
Accounting only for carrier-light interaction, the applied ultrashort pulse excites electrons from the valence into the conduction band resulting in an optically induced carrier density $\mathcal{N}_{\rm{opt}}(t)$,  which remains constant after the pulse is switched off (green line). Taking into account Coulomb-induced scattering channels including Auger processes, we find a strong increase of the carrier density even after the optical pulse has been switched off (red line) - carrier multiplication takes place.  In the case of a purely Coulomb-driven dynamics, we obtain a multiplication factor of $\rm{CM}=3.7$. However, the dynamics is still incomplete, as phonon-induced relaxation channels have not been included yet. Phonons directly compete with Auger channels for excited carriers that can scatter to an energetically lower state by either performing an Auger process or by emitting a phonon and thus reducing the CM efficiency.   In spite of this, taking the full dynamics into account including carrier-phonon relaxation channels, we still observe a significant carrier multiplication with a maximum value of approximately $\rm{CM}=2.5$ that persists on a time scale of a few picoseconds  \cite{winzer10,winzer12b}, cf. the blue line in Fig. \ref{fig4_cm}.

The theoretically predicted carrier multiplication in graphene has been recently demonstrated in high - resolution multi-color pump-probe \cite{ploetzing14,brida13} and time-and angle-resolved
photoemission (ARPES) measurements \cite{gierz15}.
Gierz and co-workers \cite{gierz15}  applied extreme-ultraviolet pulses  to track the number of excited electrons and their kinetic energy in ARPES measurements. In a time window of approximately \unit[25]{fs} after the
pump pulse,  a clear increase of carrier density and a simultaneous decrease of the
average carrier kinetic energy was observed directly revealing that relaxation is dominated by impact excitation \cite{gierz15}.
Furthermore, 
as absorption at optical frequencies is dominated by
interband processes, pump-probe experiments can be also used to 
directly monitor carrier occupation probabilities of the
optically coupled states as well as the Coulomb- and phonon-induced
carrier escape from these states. We have  performed a series of pump-probe measurements \cite{ploetzing14} applying different optical probe energies ranging from 0.73 to \unit[1.6]{eV}, while the pump pulse was fixed at \unit[1.6]{eV}. Assuming quasi-instantaneous thermalization of the excited carriers through ultrafast carrier-carrier scattering, the measured occupation probabilities at different distinct energies could be exploited to reconstruct the time-dependent
carrier distribution in the relevant range in the momentum space \cite{ploetzing14}. To be able to draw qualitative conclusions on the appearing carrier multiplication, the optically excited carrier density ${\mathcal{N}}_{\rm{opt}}$ was estimated based on the pump fluence and the dark absorption of graphene on sapphire. We find that depending on the pump fluence, the integrated carrier density $\mathcal{N}$ can lie above ${\mathcal{N}}_{\rm{opt}}$ presenting a clear experimental prove of the appearance of CM in graphene.

\begin{figure}[t!]%
\centering
\includegraphics*[width=0.75\columnwidth]{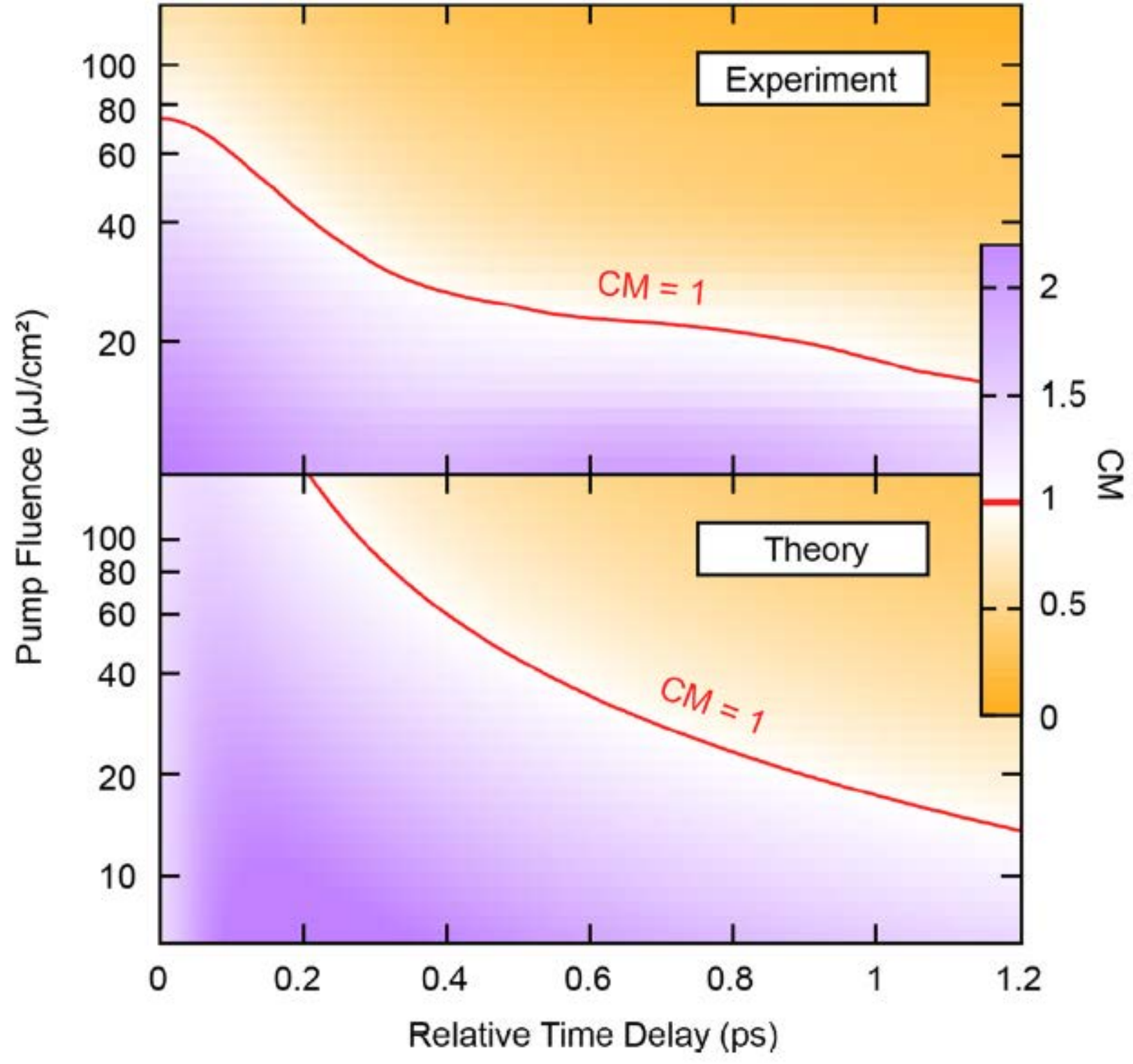}
\caption{\textbf{Experiment-theory comparison on carrier multiplication.} Theoretically predicted and experimentally measured carrier multiplication as  a function of time and pump fluence. Purple colors mark the region, where carrier multiplication takes place, i.e. CM >1. We find both in experiment  and theory a maximum CM at low pump fluences on a timescale of a few picoseconds. Figure taken from Ref. \cite{ploetzing14}. }
\label{fig5_cm_exp_theo}
\end{figure}

Figure \ref{fig5_cm_exp_theo} shows a direct comparison between the theoretically predicted and experimentally measured carrier multiplication revealing an excellent agreement  with respect to the quantitative CM values as well as its qualitative dependence on the pump fluence \cite{ploetzing14}.  
 Depending on the pump fluence and the time after the optical pulse, we can clearly distinguish both in experiment and theory two distinct regions characterized by $\rm{CM}>1$ (purple) and $\rm{CM}<1$ (orange). 
Surprisingly, we find the largest CM values at low pump fluences in the range of $\unit[1-10]{\rm{\mu} J cm^{-2}}$ with lifetimes in the range of several picoseconds. At intermediate  fluences up to approximately $\unit[80]{\rm{\mu} J cm^{-2}}$, we observe a smaller carrier multiplication on a much shorter timescale. Here, IE still prevails over AR, however, the asymmetry between these two processes is restricted to a shorter time range, since the number of scattering partners is increased accelerating the relaxation dynamics and leading to a faster equilibration between the IE and AR processes \cite{winzer12b}. In the strong excitation regime with pump fluences larger than $\unit[100]{\rm{\mu} J cm^{-2}}$, the states are highly occupied and Pauli blocking prefers AR bringing carriers back to the valence band. As a  result, the carrier density decreases and we find a \textit{negative} carrier multiplication $\rm{CM}<1$ \cite{ploetzing14}, cf. Fig. \ref{fig5_cm_exp_theo}.

\begin{figure}[t!]%
\centering
\includegraphics*[width=\columnwidth]{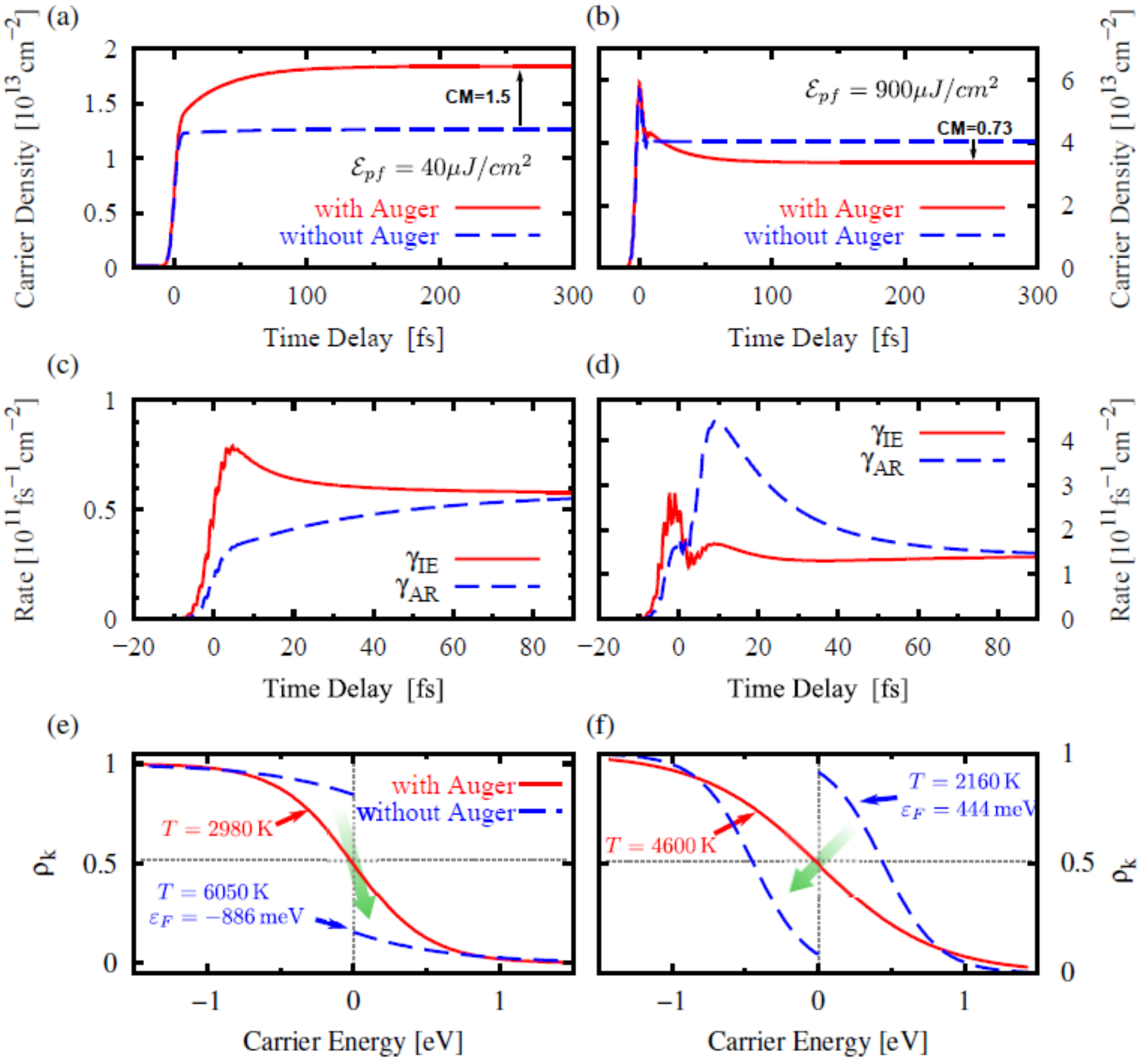}
\caption{\textbf{Fluence dependence of CM.} 
Direct comparison of the carrier density, Auger scattering rates and the spectral carrier distribution in a relatively weak (left panel) and a strong (right panel) excitation regime for exemplary pump fluences
 of $\varepsilon_{pf}= \unit[40]{\rm{\mu} J/cm^2}$ and $\unit[900]{\rm{\mu} J/cm^2}$, respectively. (a)-(b) Temporal evolution of the carrier density accounting for the Coulomb-driven carrier dynamics with (red solid line) and without  Auger
processes (blue dashed line). (c)-(d)  Temporal evolution of Auger generation (IE) and recombination rates (AR).  (e)-(f) Spectral carrier distribution after accomplished Coulomb-driven dynamics again with and without Auger processes. Efficient Auger scattering drives the electronic system to an equilibrium distribution characterized by a zero Fermi energy $\varepsilon_F=\unit[0]{eV}$. In absence of these channels, a negative or positive Fermi energy is obtained depending on the excitation regime. 
Figure taken from Ref. \cite{torben_phd}.
}
\label{fig6_cm_fluence}
\end{figure}

To get deeper insights into the influence of  the excitation strength on the carrier multiplication, we show  the temporal evolution of the carrier density and Auger rates as well as the final spectral carrier distribution at two exemplary pump fluences of $\unit[40]{\rm{\mu} J/cm^2}$ and $\unit[900]{\rm{\mu} J/cm^2}$, cf. Fig. \ref{fig6_cm_fluence}. 
Calculating the carrier density accounting for all carrier-carrier scattering channels (solid red line) and excluding Auger processes (blue dashed line), we directly demonstrate the impact of the latter. 
At $\unit[40]{\rm{\mu} J/cm^2}$, we still find a carrier multiplication (Fig. \ref{fig6_cm_fluence}(a)) , however with a reduced factor of 1.5 compared to Fig. \ref{fig4_cm}, where a pump fluence of $\unit[2.3]{\rm{\mu} J/cm^2}$ has been assumed. At $\unit[900]{\rm{\mu} J/cm^2}$, Auger processes even account for a lower overall carrier density resulting in CM < 1 (Fig. \ref{fig6_cm_fluence}(b)). To understand this behavior, we study the time-dependent scattering rates for the impact excitation (IE) and the Auger recombination (AR), cf. Fig. \ref{fig6_cm_fluence}(c)-(d). At $\unit[40]{\rm{\mu} J/cm^2}$, we observe a clear asymmetry between
both rates in favor of the impact excitation in the first \unit[100]{fs}. This represents the time window for carrier multiplication. We realize that this time range is considerably shorter comparing with Fig. \ref{fig4_cm} where CM takes place in the first \unit[300]{fs}. At the much higher pump fluence of $\unit[900]{\rm{\mu} J/cm^2}$, we find that the imbalance between the Auger rates
still initially favors IE. However, approximately at the center of the excitation a transition can be observed and the Auger recombination rate exceeds the impact excitation. 
 Auger processes give rise to an efficient carrier recombination until an equilibrium distribution is reached. This results in a reduction of optically generated carriers that is reflected by  CM = 0.73. 

Since the corresponding Coulomb matrix elements are the same for IE and AR, the observed asymmetry can be ascribed to Pauli blocking effects.
To get further insights, we compare the spectral distribution of carriers after the Coulomb-driven relaxation dynamics with
and without Auger processes (Fig. \ref{fig6_cm_fluence}(e)-(f)). We find that in both excitation regimes Auger processes  lead
to a single Fermi-Dirac distribution  characterized by a zero Fermi energy.
In contrast, neglecting Auger processes leads to distinct distributions in
conduction and valence band with finite Fermi levels, respectively. In the case of the weaker excitation, the Fermi energy is negative. To 
achieve an equilibrium carrier distribution, the generation of carriers via impact excitation is required, i.e. CM > 1.  At the strong excitation the Fermi energy is positive
and only predominant processes of  Auger recombination can prevent the discontinuity in the carrier distribution between both bands resulting in CM < 1. 
The imbalance between impact excitation and Auger recombination can be also explained in terms of Pauli blocking:
In the case of weak excitation, the valence band occupation close to the Dirac point is large compared to the respective conduction band occupation, which gives rise to efficient impact excitation,
cf. green arrow in Fig. \ref{fig6_cm_fluence}(e). Forstrong excitation the population conditions around the Dirac point are reversed quickly after the optical excitation, and thus Pauli blocking favors Auger recombination (Fig. \ref{fig6_cm_fluence}(f)).

\subsection{Carrier multiplication in Landau-quantized graphene}
While the carrier dynamics in graphene
has been intensively investigated over the last years,
there have been only a few studies on the dynamics in Landau-quantized graphene \cite{plochocka09, wendler13, wendler14,mit14, wang14, winnerl15, wendler16}.
The first experiment  was performed by Plochocka et al. in
2009 \cite{plochocka09}, where rather high-energetic Landau
levels ($n\sim100$) have been investigated. A strong suppression of Auger processes was observed and explained
as a consequence of the non-equidistant level spacing.
However, due to the $E \propto \sqrt{n}$ dependence,  there are Landau levels $n$ that are actually equidistant suggesting that Auger scattering might be important in certain situations. 
Here, we suggest a specific pumping scheme to open up Auger scattering channels and to achieve a carrier multiplication in Landau-quantized graphene. The strategy is to excite
charge carriers to $n=+4$, which induces an energy conserving
scattering process consisting of $+4\rightarrow  +1$
and $0\rightarrow +1$. This relaxation channel corresponds to impact excitation and creates additional
charge carriers. Due to Pauli blocking, the inverse process of Auger recombination is expected to be suppressed. 

\begin{figure*}[t!]%
\centering
\includegraphics*[width=\textwidth]{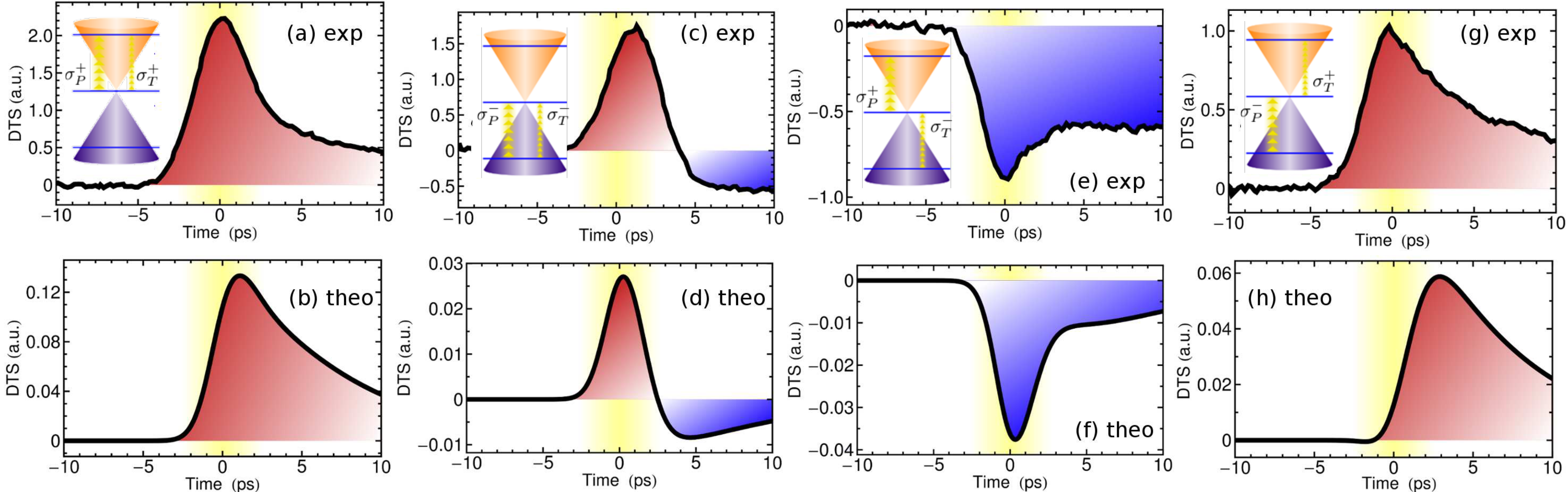}
\caption{\textbf{Experiment-theory comparison on Auger scattering in Landau-quantized graphene.} Direct comparison of experimentally measured (upper panel) and theoretically predicted (lower panel) polarization-resolved differential transmission for pumping and probing the same transition [(a)-(d)] and different transitions [(e)-(h)]. The corresponding pump ($\sigma_{\text{P}}$) and probe (or test $\sigma_{\text{T}}$) pulses are sketched in the insets. The yellow shaded areas in the background illustrate the width of the pump pulse. 
Figure adapted from Ref. \cite{winnerl15}.}
\label{fig7_cm_LL_exp}
\end{figure*}

Applying an optical pulse with an excitation energy of $\unit[280]{meV}$ matching the transitions 
$\mp 4 \rightleftarrows \pm 3$
at a reasonable magnetic field of $B=\unit[4]{T}$, we find an efficient impact excitation even in Landau-quantized graphene resulting in a carrier multiplication \cite{wendler14}.
The theoretically predicted importance of Auger scattering in Landau-quantized graphene is surprising in view of the  non-equidistant Landau level separation.  
Since  the carrier occupation of single LLs can be directly addressed in polarization-dependent pump-probe experiments, we can track the way of excited carriers in specific Landau levels and can experimentally investigate the importance of Auger channels. We apply circularly polarized light of a specific energy to selectively pump and probe transitions
between the energetically lowest Landau levels $n=  -1, 0$, 
and $ +1$. We perform four different experiments pumping and probing the same and different LL transitions, respectively, cf. Fig. \ref{fig7_cm_LL_exp}.
Considering only the optical excitation and the Pauli blocking, we expect a positive differential transmission signal (DTS), if the pump $\sigma_P$ and the probe (test) pulse $\sigma_T$ have the same polarization.  Here, the excitation of charge carriers due to the pump pulse is expected to lead to an absorption bleaching of the probe pulse due to the increased Pauli blocking. Reduction in absorption directly translates in an increase of differential transmission. 
 In contrast, using an opposite polarization
for the pump and the probe pulse, we either depopulate the final state or populate the initial state for the absorption of the probe pulse. As a result, Pauli blocking is expected to be reduced giving rise to an absorption enhancement and a negative DTS.

Figure \ref{fig7_cm_LL_exp} illustrates a direct comparison between  experimentally measured (upper panel)  and theoretical predicted (lower panel) differential transmission spectra for the four configurations of pump and probe
pulse polarizations. Both pulses have the same width of $\sigma_{t}=\unit[2.7]{ps}$ (yellow area in Fig. \ref{fig7_cm_LL_exp}) and the same energy of $\unit[75]{meV}$. Comparing the obtained results with
the above expectation only based on the occupation change induced by the optical
excitation and Pauli blocking, we find the expected positive and negative DTS signs in the case of  pumping with  $\sigma_{\text{P}}^{+}$-polarized light, cf. Fig. \ref{fig7_cm_LL_exp})(a)-(b) and (e)-(f). However, we reveal a qualitative difference both in experiment and theory in the case of pumping with  $\sigma_{\text{P}}^{-}$-polarized light: (i) The configuration $\sigma_{\text{P}}^{-},\,\sigma_{\text{T}}^{-}$  (Figure \ref{fig7_cm_LL_exp}(c)-(d)) shows an initial expected increase in DTS that is however followed by an unexpected DTS sign change. (ii) The configuration $\sigma_{\text{P}}^{-},\,\sigma_{\text{T}}^{+}$ (Figure \ref{fig7_cm_LL_exp}(g)-(h)) shows a completely contrary behavior. We find an entirely positive DTS, although the pump pulse populates the zeroth Landau level that is the initial state for the absorption of the probe pulse, i.e. we would expect an absorption enhancement and a negative DTS.

To explain these surprising results, we need to go beyond  simple expectations based on Pauli blocking. We have to take into account the full carrier dynamics, in particular including Auger scattering processes. Furthermore, we need to include doping to break the electron-hole symmetry to be able to reproduce the experimental observations. Otherwise, in an undoped system, the two configurations $\sigma_{\text{P}}^{+},\,\sigma_{\text{T}}^{+}$
and $\sigma_{\text{P}}^{-},\,\sigma_{\text{T}}^{-}$ (and likewise
$\sigma_{\text{P}}^{+},\,\sigma_{\text{T}}^{-}$ and $\sigma_{\text{P}}^{-},\,\sigma_{\text{T}}^{+}$)
would yield the exactly same DTS. The assumption of a finite doping
is supported by experimental studies showing that multilayer
epitaxial graphene samples grown on the C-terminated face of SiC have a finite n-doping due to
a charge transfer from the SiC substrate \cite{sprinkle09,sun10b}.
Therefore, for the theoretical calculations a Fermi energy of $\varepsilon_{\text{F}}=\unit[28]{meV}$ has been assumed.

\begin{figure}[t!]%
\centering
\includegraphics[width=\columnwidth]{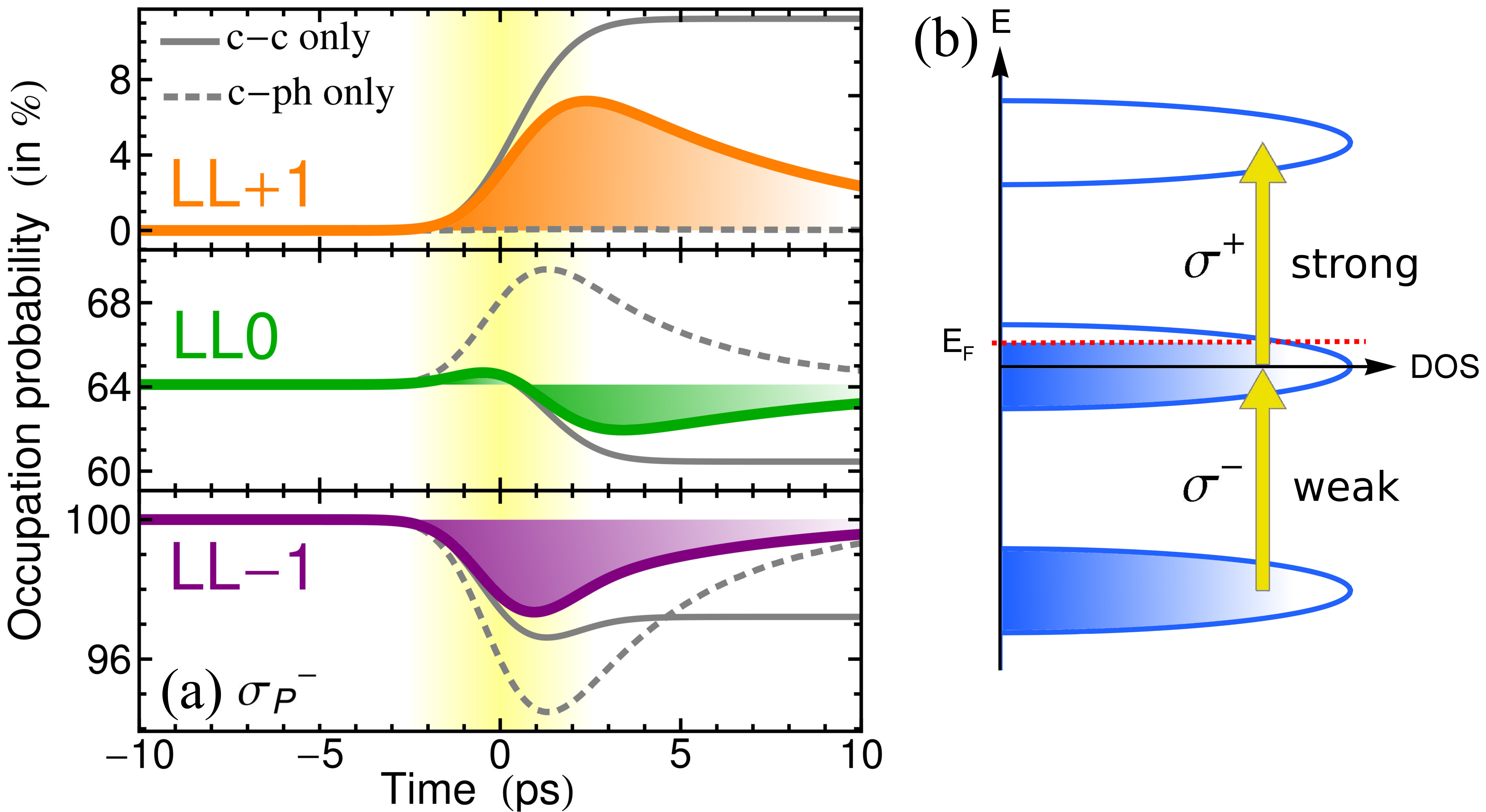}
\caption{\textbf{Microscopic view on impact of Auger processes.}  (a) Temporal evolution of the carrier occupation in the three energetically lowest LLs that are  involved in the Auger processes from Figure \ref{fig7_cm_LL_exp}
after excitation with $\sigma^{-}$-polarized light. Solid and dashed gray lines indicate the dynamics accounting only for carrier-carrier and carrier-phonon processes, respectively. 
In presence of Auger processes, LL0 becomes depopulated despite the optical pumping into this state. (b) In the case of n-doping (positive Fermi energy), the efficiency of optical pumping is reduced for
$\sigma^{-}$-polarized light due to Pauli blocking.
 Figure (a) adapted from Ref. \cite{winnerl15} and (b) from Ref. \cite{florian_phd}.}
\label{fig8_cm_LL_exp_sol}
\end{figure}

Solving the Graphene Bloch Equations, we have all tools at hand to provide a microscopic view on the carrier dynamics in Landau-quantized graphene and to resolve the unexpected DTS sign observed in Figure \ref{fig7_cm_LL_exp}(g). To get the required insights, we calculate the temporal evolution of the involved LL occupations $\rho_{+1}$, $\rho_{0}$ and $\rho_{-1}$
after the application of $\sigma^{-}$-polarized pump pulse, cf. Fig. \ref{fig8_cm_LL_exp_sol}. To identify the role of  Auger scattering, we compare the full dynamics with a calculation neglecting Auger processes (dashed gray lines). Note that  carrier-phonon scattering has been phenomenologically included to match the experimentally observed fast decay rates at long times.  Surprisingly, our calculations reveal that  the carrier occupation of the zeroth LL $\rho_{0}$  shows only a minor increase in the beginning of the dynamics and it  even starts to decrease already before the center of the pulse is reached. This means that although we optically pump carriers into the zeroth Landau level, its population actually decreases. This surprising result can be explained by extremely efficient Auger scattering, which  induces the
transitions $ 0\rightarrow -1$ and $ 0\rightarrow +1$ resulting in a quick depopulation of $\rho_{0}$, cf. Figure \ref{fig8_cm_LL_exp_sol}(b). The crucial role of the Auger scattering becomes apparent by comparing the temporal evolution of $\rho_{0}$ to the case without Auger processes (dashed gray lines). Here, as expected, the population of the zeroth Landau level increases during the entire time of the optical excitation. This remarkable many-particle effect appears only in the case of optical pumping with  $\sigma_{\text{P}}^{-}$-polarized light, since here the pumping efficiency is strongly reduced
due to an enhanced Pauli blocking as a result of a finite n-doping, cf. Fig. \ref{fig8_cm_LL_exp_sol}(b).

To sum up, the polarization-dependent pump-probe experiments provide
a clear evidence for an efficient Auger scattering in Landau-quantized graphene. The observed unexpected DTS sign  emerges, since Auger scattering depopulates the zeroth Landau level faster than it is filled by optical excitation.

\section{Population inversion}
Population inversion (PI) is a many-particle phenomenon, where the thermal carrier distribution is inverted, i.e. energetically higher states  become stronger occupied than energetically lower states. This configuration can lead to light amplification and optical gain that  are of central importance in optical sciences \cite{haken84,haken95,scully97}. They are based on the quantum effect of induced photon emission and constitute one of the key ingredients for laser technology and optical data communication \cite{scully97, winzer13, jago15, brem16}. 
Visible and infrared fiber lasers are the basis for information technology, while microwave and radio-frequency emitters build the backbone of wireless communications. Due to the lack of efficient sources of terahertz light, there is a technological gap between these two frequency ranges  \cite{koehler02}. 

The ongoing search for novel gain materials has brought graphene into the focus of research. Its linear and gapless band structure  offers a broad spectrum of optically active states including the terahertz region. This extraordinary feature has already been technologically exploited in graphene-based photodetectors covering a wide range of frequencies \cite{xia09, mueller10, echtermeyer11, furchi12} as well as graphene-based saturable absorbers converting the continuous wave output of lasers into a train of ultrashort optical pulses \cite{bonaccorso10, avouris12}.  A spectrally broad population inversion has been measured \cite{li12,gierz13} and theoretically predicted \cite{ryzhii07,winzer13, jago15} in the strong excitation regime. Recently, a population inversion in Landau-quantized graphene has also been predicted \cite{wendler15, belyanin15, brem16} suggesting the design of tunable terhertz Landau level lasers.

\subsection{Population inversion in graphene}

\begin{figure}[t!]
  \begin{center}
\includegraphics[width=0.7\linewidth]{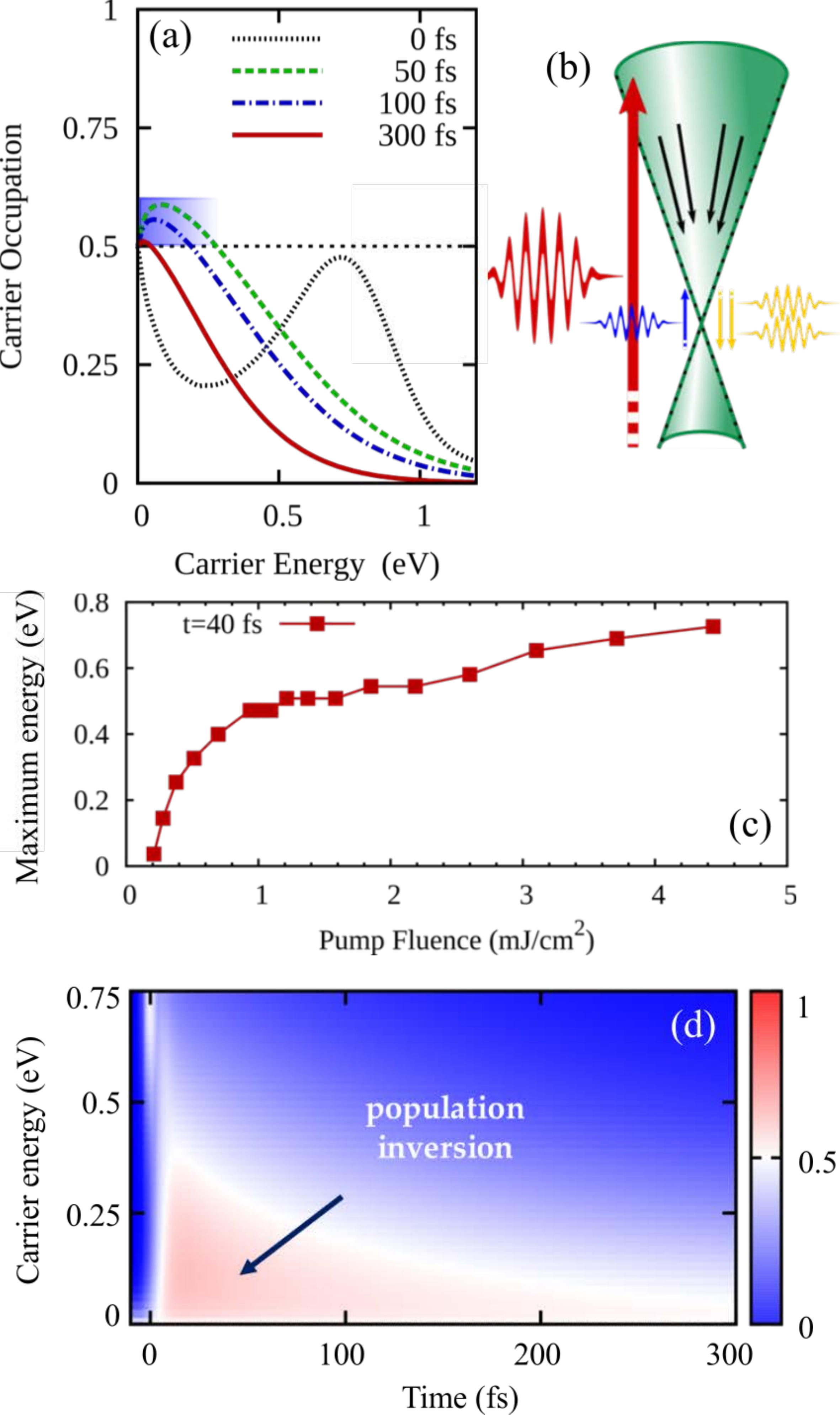}
  \end{center}
  \caption{\textbf{Population inversion.} (a) Spectrally resolved carrier occupation $\rho^c_{\bf k}(t)$ in the conduction band at different times after an optical excitation with a pump fluence of approximately $\unit[2.5]{mJ/cm^2}$. We observe a population inversion, i.e. $\rho^c_{\bf k}(t)>0.5$, during the first $\unit[300]{fs}$ for carrier energies smaller than $\unit[250]{meV}$.  (b) Scheme of optical pumping (red) and the build-up of a phonon bottleneck (black). Here, an energetically lower probe pulse (blue) can be amplified via induced emission (yellow), if population inversion occurs.  
(c) Maximal spectral region  up to which a population inversion can be achieved as a function of the pump fluence. The values are extracted from the carrier distribution $\unit[40]{fs}$ after the excitation pulse.
(d) Spectrally and temporally resolved carrier occupation in the conduction band demonstrating generation and decay of population inversion at a pump fluence of $\unit[2.5]{mJ/cm^2}$.
 Figures (a)-(c) adapted from Ref. \cite{winzer13} and (d) from Ref. \cite{torben_phd}. }\label{fig9_PI}  
\end{figure}

Based on Graphene Bloch Equations, we investigate the carrier dynamics in the strong excitation regime addressing the question whether a population inversion can be achieved in graphene. We apply an optical pulse with a pump fluence of $\unit[2.5]{mJ/cm^2}$ that is three orders of magnitude higher than in Fig. \ref{fig3_dynamics}. The carrier occupation in the conduction band is illustrated in Fig. \ref{fig9_PI}(a) as a function carrier energy for different times after the strong optical excitation. At \unit[0]{fs} corresponding to the maximum of the excitation pulse, we find a strongly pronounced non-equilibrium distribution that is centered at the carrier energy of \unit[0.75]{eV} reflecting to the excitation energy of \unit[1.5]{eV}. We observe how the excited carriers become quickly distributed leading to a spectrally broadened distribution. Interestingly, in contrast to the dynamics in the low excitation regime (Figure \ref{fig3_dynamics}) we do not observe a thermalized Fermi distribution, but find a rather surprising feature at low energies: the carrier occupation $\rho_{\bf k}^c$ exceeds the value of 0.5 in a spectrally and temporally limited region (blue shaded area). This means that the occupation of the energetically higher states in the conduction band is actually higher than the occupation in the corresponding states in the valence band. We observe this population inversion for carrier energies up to \unit[250]{meV} and times up to \unit[300]{fs} after the optical excitation, cf. Figure \ref{fig9_PI}(a). This finding is in excellent
agreement with a recent experimental study demonstrating the appearance of a temporally and spectrally
limited optical gain in graphene \cite{li12}. The achieved PI values as well as its spectral and temporal region strongly depend on the applied pump fluence.
Figure \ref{fig9_PI}(c) illustrates the fluence-dependent maximal energy, up to  which PI can be obtained. The values are taken $\unit[40]{fs}$ after the excitation  pulse corresponding to the temporal resolution of the experiment \cite{li12}. Population inversion appears above a threshold pump fluence of approximately $\unit[200]{\rm{\mu} J/cm^2}$, where it is limited to very small energies. For stronger excitations, the PI regime is broadened up to $\unit[0.75]{eV}$. Finally, Fig. \ref{fig9_PI}(d) illustrates the build-up and the decay of the population inversion at a pump fluence of approximately $\unit[2.5]{mJ/cm^2}$. The surface plot shows the carrier occupation $\rho^c_{\bf k}(t)$ as a function of time and energy. We find a distinct region with $\rho^c_{\bf k}(t)>0.5$ (red area), where PI takes place. Our calculations reveal that the PI is generated in a broad spectral range almost instantaneously in the first femtoseconds after the optical excitation. As discussed above, this can be ascribed to the ultrafast carrier-phonon intraband scattering combined with a vanishing density of states around the Dirac point.  Already after a few femtoseconds, the PI maximum is reached that is followed by a Auger-induced decay on a time scale of a few hundreds of femtoseconds.

Exploiting the microscopic access to time- and momentum - resolved carrier dynamics, we can reveal the microscopic mechanism behind the generation of population inversion in graphene (Figure \ref{fig9_PI}(b)) \cite{winzer13}:
 (i) Strong optical excitation (red arrow) lifts many electrons from the valence into the conduction band. (ii) Intraband scattering processes (black arrows) bring excited carriers in the vicinity of the Dirac point and due to the vanishingly small density of states in this region,  carriers accumulate and above a certain threshold pump fluence a population inversion appears. (iii) Stimulated emission (yellow arrows) amplifies the probe 
pulse (blue arrow) at low energies. 
Our calculations clearly demonstrate that intraband scattering with optical phonons is the crucial relaxation channel for the build-up of the PI. The gain regime is strongly reduced, if we consider only Coulomb-induced intraband processes. In particular, impact excitation plays a minor role, since in the strong excitation regime, the inverse process of Auger recombination is more important. It reduces the accumulation of carriers close to the Dirac point and leads to a decay of the PI on a time scale of few hundreds femtoseconds. Phonon-induced recombination are an order of magnitude weaker than AR processes \cite{winzer13}.

So far, we have shown that a transient population inversion can be achieved in graphene. However, a long-lived
optical gain is the key prerequisite for the realization of graphene-based laser devices that could also operate in the technologically relevant terahertz spectral region. In a recent study, we have proposed a strategy
of how to achieve long-lived gain \cite{jago15}: 
(i) reduction of the efficiency of Auger
recombination by studying graphene on a high-dielectric
substrate and at the same time (i) enhancement of the carrier-light
interaction by integrating graphene into a high-quality
photonic crystal nanocavity. Following this recipe, we have shown that  coherent laser
light emission can be achieved  from graphene \cite{jago15}.

\subsection{Population inversion in Landau-quantized graphene}

The idea of a two-dimensional Landau level
laser dates back to 1986, when H. Aoki proposed to exploit the energetic LL spacing in two-dimensional electron gases to externally tune the laser frequency \cite{aoki86}. The key challenge for the realization of such a laser is a stable population inversion. However, in conventional semiconductors exhibiting an equidistant spectrum of LLs, strong Coulomb scattering acts in favor of an equilibrium Fermi-Dirac distribution and strongly counteracts the build-up of a population inversion. 
 In contrast, graphene exhibits a non-equidistant LL separation and specific optical selection rules offering optimal conditions
to overcome the counteracting Coulomb- and phonon-assisted scattering channels and to achieve long-lived population inversion \cite{aoki09,wendler15b,belyanin15}.

\begin{figure}[t!]
\begin{centering}
\includegraphics[width=.75\linewidth]{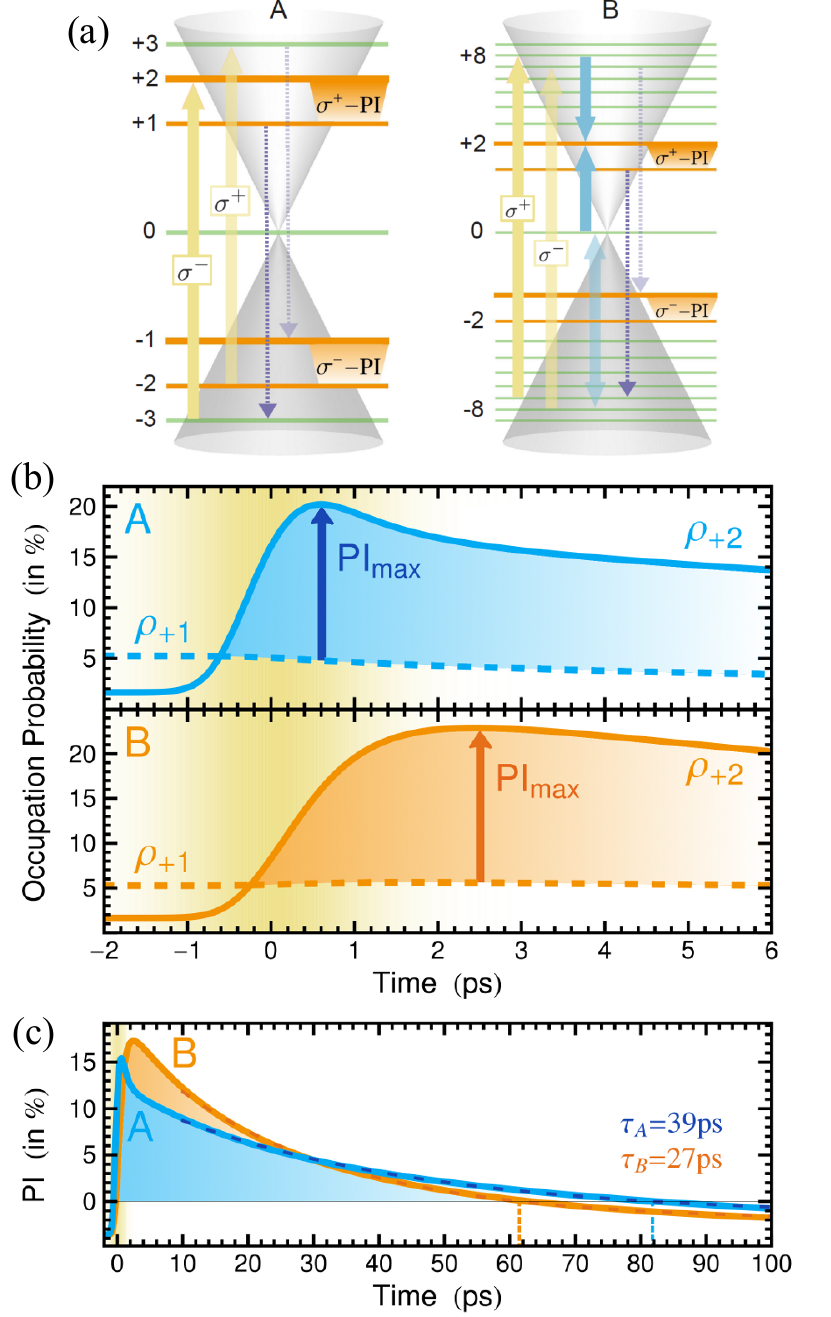}
\par\end{centering}
\caption{\textbf{Population inversion in Landau-quantized graphene.} 
(a) Schematic illustration of two different schemes to achieve population inversion between the LLs $n=+1$
and $n=+2$. Scheme A is
based only on optical pumping exploiting the selection rules (yellow arrows), while scheme B is based on efficient Auger scattering (blue arrows)
between the equidistant LLs $n=+8,+2,$ and $0$ redistributing
optically pumped electrons from $n= +8$ to $n= +2$. To allow for continuous laser action,
phonon-assisted relaxation channels are needed  connecting all LLs involved in the laser system
(purple-dotted arrows). 
(b) Temporal evolution of the carrier occupations $\rho_{+1}$
and $\rho_{+2}$ for both PI schemes. It illustrates the appearance of a pronounced population inversion defined by
$\text{PI}=\rho_{+2}-\rho_{+1}>0$
(blue and orange shaded areas). 
(c) Temporal evolution of the PI characterized by an ultrafast build-up and a
slow decay on a picosecond time scale. Figure taken from Ref. \cite{wendler15b}.}
\label{fig10_PI_lqg}
\end{figure}

Based on our microscopic insights into  the underlying many-particle mechanisms, we propose two different experimentally realizable schemes to design tunable graphene-based THz Landau level lasers (Fig. \ref{fig10_PI_lqg}(a)): (A) The first scheme is only based on optical pumping exploiting the specific optical selection rules in Landau-quantized
graphene allowing selective pumping of a single LL. (B) The second scheme exploits  efficient Auger scattering to assist the build-up of a PI. An Auger-induced mechanism to create P is quite
remarkable, since  Auger scattering is known to rather be the main obstacle for the realization of population inversion \cite{winzer13, plochocka09,gierz15b}. 

Applying linearly polarized light, we preserve the electron-hole symmetry. As a result, we obtain a PI both in the conduction band and in the valence band. In the considered scheme, it occurs between the LLs with the indices
$n=\pm1$ and $n=\pm2$.  Note that the PI transitions in the conduction and in the valence band are optically
coupled by inversely circularly polarized photons, i.e. photons created
in a stimulated emission process inducing the LL transition
$+2\rightarrow +1 (-1\rightarrow -2$)
are $\sigma^{+}$- polarized ($\sigma^{-}$- polarized). Hence, we
label the corresponding population inversion as $\sigma^{+}$- PI
and $\sigma^{-}$- PI, respectively.

Now, we investigate the applicability of the two proposed schemes for population inversion. We evaluate the Graphene Bloch Equations in the presence of an external magnetic field $B=\unit[4]{T}$. We optically excited the system by applying a pump pulse with a width of $\unit[1]{ps}$, a pump fluence of $\varepsilon_{\text{pf}}=\unit[1]{\rm{\mu} Jcm^{-2}}$,
and an energy matching the pumped LL transitions of the respective PI scheme. 
To have a fair comparison, we keep the excitation energy-dependent pulse area constant and increase the pump fluence in scheme B to $\varepsilon_{\text{pf}}=\unit[2.27]{\rm{\mu} Jcm^{-2}}$.
Since undoped Landau-quantized graphene is symmetric for electrons and holes, we
focus on the discussion of the $\sigma^{+}$-PI in the conduction band.

Figures \ref{fig10_PI_lqg}(b)-(c) illustrates the time-dependent occupations 
$\rho_{+1}$ and $\rho_{+2}$ and
the resulting population inversion, respectively. 
In both PI schemes, $\rho_{+1}(t)$ remains nearly constant, while $\rho_{+2}(t)$ strongly increases on the
time scale of the optical excitation (yellow area in Fig. \ref{fig10_PI_lqg}(b)) followed by a slow decay. We observe a long-lived population inversion between the two Landau levels. The PI defined as  $\text{PI}=\rho_{+2}-\rho_{+1}>0$ is illustrated in Fig. \ref{fig10_PI_lqg}(c). Exponential fits reveal a slower decay time in the A scheme ($\tau_{A}=\unit[39]{ps}$  vs. $\tau_{B}=\unit[27]{ps}$).
 In scheme B,  Auger processes are required
to induce population inversion resulting in its delayed build-up. The PI maximum is reached a few picoseconds after the excitation pulse. While both PI schemes
are suited to create a significant PI with a rather long decay time,
the advantage of scheme B is its additional Landau level making it
a potentially better laser system, which is reflected by its higher
maximal PI value.

The predicted long-lived population inversion is a crucial step towards the design of graphene-based Landau level lasers. 
To address the question whether such a laser can work under realistic experimental conditions, we have developed a fully quantum mechanical theoretical approach going beyond the Graphene Bloch Equation by explicitly including the dynamics of photons. This way, we obtain microscopic access to the coupled dynamics of electrons, phonons, and photons in Landau-quantized graphene. We show that embedding graphene  into a high quality Fabry-Perot microcavity \cite{krupke12} with a resonator mode matching the LL transition $+1\rightarrow +2$, the trapped cavity photons become multiplied via stimulated emission resulting indeed in an emission of terahertz laser light. To achieve continuous wave lasing, where carriers perform cycles in a three-level system, we show that the lasing process  needs to be complemented by  emission of optical phonons. The latter depopulate the lower laser level $n=+1$  and repopulate the initial LL $n=-3$  for optical excitation. Our calculations reveal that for a high-quality cavity and sufficient pump power, the
laser frequency can be externally tuned within a continuous range between 4 and \unit[8.5]{THz} \cite{brem16}. \\

In summary, we have presented a review of our joint theory-experiment research on 
ultrafast carrier dynamics in graphene and  Landau-quantized graphene. 
We provide a microscopic view on elementary Coulomb- and
phonon-induced scattering processes characterizing the non-equilibrium
carrier dynamics as well as provide a direct comparison with recent experimental observations. In particular, we focus on two technologically promising ultrafast phenomena characterizing the carrier dynamics in graphene: carrier multiplication and population inversion. We find that due to graphene's linear and gapless electronic band structure, Auger scattering processes bridging the valence and the conduction band dominate the carrier dynamics both in graphene and Landau-quantized graphene resulting in a significant multiplication of optically excited charge carriers. 
Furthermore, we demonstrate the appearance of a population inversion and discuss how this many-particle phenomenon can be exploited to design externally tunable Landau level lasers that are operating in the technologically relevant terahertz spectral range. \\

We are very grateful to the German Research Foundation (DFG) for organizing the Priority Program (SPP) 1459 Graphene and to Thomas Seyller (Chemnitz University of Technology) for the coordination of the program. Our work 
has also received funding from the European Union's Horizon 2020 research and innovation programme
under grant agreement No 696656 (Graphene Flagship) and the Swedish Research Council (VR).

\section*{Appendix}

Here, we describe the applied theoretical approach in more detail. Note that the appendix is based on Refs. \cite{malic11b, malic13}.

\paragraph*{Hamilton operator:}  The first step is the description of the many-particle Hamilton operator $H$ in 
the formalism of second quantization. To this end, we introduce ladder  operators $a_{\bm l}^+$ and
$a^{\phantom{+}}_{\bm l}$ ($b_{\bm u}^+$,
$b^{\phantom{+}}_{\bm u}$) creating and annihilating an electron (phonon) in the state $\bm l$ ($\bm u$), respectively. 
The compound index $\bm {l}=(\lambda,\bm {k})$ contains
 the electronic momentum $\bm k$ and the conduction/valence band $\lambda=v,c$, while $\bm {u}=(j,\bm {q})$ describes the
phonon momentum $\bm q$ and the phonon mode $j$. In our work, the many-particle Hamilton operator $H=H_{0}+H_{c,l}+H_{c,c}+H_{c,p}$ consists of the interaction-free carrier and phonon part $H_0$, the carrier-light $H_{c,f}$, the carrier-carrier $H_{c,c}$, and the carrier-phonon  interaction $H_{c,p}$.  

The first part $H_0$ is determined by the electron ($\varepsilon_{\bm l}$) and the phonon dispersion ($\hbar \omega_{\bm u}$):
\begin{equation}
\label{h_free}
 H_{0} = \sum_{\bm{l}}\varepsilon_{\bm{l}}^{}a_{\bm{l}}^{+}a_{\bm{l}}^{\phantom{+}} +\sum_{\bm{u}}\hbar\omega_{\bm{u}}^{}(b_{\bm{u}}^{+}b_{\bm{u}}^{\phantom{+}}+\frac{1}{2})\,.
\end{equation}
The electronic dispersion is calculated analytically within
the nearest-neighbor tight-binding (TB) approach  by introducing electronic wave functions $\Psi_{\lambda \bm{k}}(\bm{r})$ approximated as a linear combination of the atomic orbital functions \cite{reich02b, malic13,malic10b}.
The appearing tight-binding overlap $\gamma_0$ determines the slope of the Dirac bands and is fixed to the experimentally measured value.  An improved tight-binding electronic dispersion can be achieved by including
 third-nearest-neighbor interactions and their overlaps \cite{reich02b} or by including the influence of energetically higher $\sigma$-bands \cite{jiang07}. However, since we focus on the carrier relaxation dynamics close to the Dirac point, the nearest-neighbor TB approximation is sufficient. 
 Regarding the dispersion of phonons, we assume constant optical phonon energies around the high-symmetry points exhibiting strong electron-phonon coupling  ($\Gamma$ and K longitudinal and transversal optical modes) \cite{piscanec04, maultzsch04}. 
 For acoustic phonons, we assume a linear phonon dispersion  taking into account the strongest $\Gamma$-LA phonon mode \cite{tse09}.

The carrier-light interaction $H_{c,l}$ is taken within the radiation gauge
and the dipole approximation \cite{scully97}
\begin{eqnarray}
\label{cf}
  H_{c,l} = i\hbar\frac{e_{0}}{m_{0}}\sum_{\bm{l}_{1},\bm{l}_{2}}\boldsymbol{M}_{\bm{l}_{1},\bm{l}_{2}}\cdot\boldsymbol{A}(t)\,a_{\bm{l}_{1}}^{+}a_{\bm{l}_{2}}^{}
\end{eqnarray}
with the elementary charge $e_0$ and the electron mass $m_0$.
The strength of the coupling is given by
the product of the vector potential $\bm {A}(t)$ and the optical matrix element \cite{grueneis03,malic06} $\bm{M}_{\bm 
  l_1,\bm  l_2}=\int d\bm{r}\Psi_{\bm l_1}^{\ast}(\bm{r})\nabla\Psi_{\bm l_2}(\bm{r})$, which can be  analytically evaluated within the TB approach
\begin{eqnarray}
 \label{opt}
\bm M^{\lambda\lambda'}_{\bm{k}}&=& m \sum_{i=1}^3\frac{\bm b_{i}}{|\bm b_i|}\bigg(C_{\lambda\bm k}^{A*} C_{\lambda'\bm k}^B
e^{i\bm{k}\cdot \bm{b}_i} -C_{\lambda\bm k}^{B*} C_{\lambda'\bm k}^A
e^{-i\bm{k}\cdot \bm{b}_i}\bigg)
\end{eqnarray}
wit the TB coefficients $C_{\lambda \bm k}^{A/B}$. The matrix element describes the strength of the carrier-light interaction and contains optical selection rules. Since the momentum of light is negligibly small, it describes direct optical transitions with $\bm l_1= \bm l_2 = \bm k$.

The third contribution $H_{c,c}$ in the Hamilton operator  describes the carrier-carrier interaction
\begin{eqnarray}
  H_{c,c}=\frac{1}{2}\sum_{\bm{l}_{1},\bm{l}_{2},\bm{l}_{3},\bm{l}_{4}}
  V^{\bm{l}_{1}\,,\bm{l}_{2}}_{\,\bm{l}_{3},\bm{l}_{4}}
  \;a_{\bm{l}_{1}}^{+}a_{\bm{l}_{2}}^{+}a_{\bm{l}_{4}}^{\phantom{+}}a_{\bm{l}_{3}}^{\phantom{+}}
\end{eqnarray}
including the  Coulomb matrix element
$V^{\bm{l}_{1}\,,\bm{l}_{2}}_{\,\bm{l}_{3},\bm{l}_{4}}$ that  is evaluated by inserting TB wave functions with effective hydrogen $2p_z$ orbitals  \cite{hirtschulz08}
\begin{equation}
\label{coulomb}
 V^{\bm{l}_{1}\,,\bm{l}_{2}}_{\,\bm{l}_{3},\bm{l}_{4}}=V_{q}\bigg(\big(\frac{q\,a_B}{Z_{\text{eff}}}\big)^2 +1 \bigg)^{-6} g^{\bm l_1\bm l_2}_{\bm l_3,\bm l_4}\delta_{\bm q, \bm k_4-\bm k_2}\,.
\end{equation}
Here,  $a_B$ denotes the Bohr radius, $Z_{\text{eff}}$ characterizes the effective atomic number,  the Kronecker accounts for momentum conservation, and $V_q$ is the Fourier transform of the two-dimensional Coulomb potential. Furthermore, the form factor  $g^{\bm l_1,\bm l_2}_{\bm l_3,\bm l_4}$  reads
$$g^{\bm l_1,\bm l_2}_{\bm l_3,\bm l_4}=\frac{1}{4}\bigg(1+c_{\lambda_1\lambda_3}\frac{e^*(\bm k_1)e(\bm k_3)}{|e(\bm k_1)e(\bm k_3)|} \bigg)\bigg(1+c_{\lambda_2\lambda_4}\frac{e^*(\bm k_2)e(\bm k_4)}{|e(\bm k_2)e(\bm k_4)|} \bigg),$$
 where  $c_{\lambda \lambda'}$ equals +1 for intraband ($\lambda=\lambda'$) and -1 for interband processes ($\lambda\neq\lambda'$). The Coulomb matrix element prefers collinear scattering along the Dirac cone. The impact of non-collinear processes is discussed in a separate feature article in this special issue [S. Winnerl et al., citation when available] with a focus on the anisotropic carrier dynamics at very low energies. 
Due to the presence of many electrons and the surrounding material, the Coulomb interaction is screened. The effects arising from the electrons in the core states and the surrounding medium are taken into account by introducing a dielectric background constant $\varepsilon_{bg}$. The screening stemming from other valence electrons are calculated within static limit of the  Lindhard equation \cite{haug04, malic13}.

The last contribution $H_{c,p}$ of the Hamilton operator describes
the interaction between carriers and phonons \cite{kadi14}:
\begin{eqnarray}
 H_{c,p}= \sum_{\bm{l_1},\bm{l_2}}\sum_{\bm{u}} \big(g_{\bm  u}^{\bm{l_1},\bm{l_2}}\;a_{\bm{l_1}}^{+}a_{\bm{l_2}}^{\phantom{+}}b_{\bm{u}}^{\phantom{+}}+g_{\bm  u}^{\bm{l_1},\bm{l_2}}\;a_{\bm{l_2}}^{+}a_{\bm{l_1}}^{\phantom{+}}b_{-\bm{u}}^{+}\big)
\end{eqnarray}
with the  carrier-phonon matrix
elements  $g_{\bm  u}^{\bm 
  l_1,\bm l_2}$.
Piscanec and co-workers have shown that the slope of the kinks in the phonon dispersion relation is directly proportional to the square of the coupling element allowing their experimental estimation \cite{piscanec04}. 
The electron-phonon coupling elements for the $\Gamma$LO, $\Gamma$TO, and the $K$ mode read \cite{piscanec04} 
\begin{eqnarray}
\label{cp_matrix_gamma}
 |g^{\bm k\lambda\lambda'}_{\bm q \varGamma\,j}|^2&=&\frac{1}{N}\tilde{g}_\varGamma^2\left(1+c_j^{\lambda \lambda'}\cos(\varphi+\varphi')\right),\\ 
\label{cp_matrix_K}
 |g^{\bm k\lambda\lambda'}_{\bm q K}|^2&=&\frac{1}{N}\tilde{g}_K^2\left(1-c^{\lambda \lambda'}_K\cos(\varphi-\varphi')\right)
\end{eqnarray}
with $j$ denoting the $\Gamma$LO or $\Gamma$TO phonon mode. Furthermore, $\tilde{g}_K^2=0.0994$ eV$^2$ for $K$ phonons, $\tilde{g}_\varGamma^2=0.0405$ eV$^2$ for $\Gamma$LO and $\Gamma$TO phonons, and $\varphi, \varphi'$ describe the angle between the wave vectors of the involved carriers and phonons. The factor $N$ corresponds to the number of graphene unit cells. The carrier-phonon matrix elements do not depend on the momentum transfer $q$ of the involved phonons, but they exhibit a characteristic angle dependence for each phonon mode.  For interband $\Gamma$TO and $K$ phonons as well as intraband $\Gamma$LO scattering $c_j^{\lambda \lambda'}=c_K^{\lambda \lambda'}=-1$, while the behavior for intraband $\Gamma$TO and interband $\Gamma$LO and $K$ phonon scattering channels is inverse with $c_j^{\lambda \lambda'}=c_K^{\lambda \lambda'}=+1$.
For acoustic phonons, we follow the approach of Tse et al. yielding \cite{tse09} $|g^{\bm k\lambda\lambda}_{\bm q \Gamma_{LA}}|^2=\frac{1}{2N}g_{LA}^2(q)\big(1+\cos(\theta_{\bm k, \bm k- \bm q})\big)$ with $g_{LA}^2=\frac{D^2 q^2\hbar}{M L^2 \omega_q}$, where $D=16$ eV is the deformation potential, $M= \unit[7.6\cdot 10^{-8}]{gcm^{-2}}$ the graphene mass density, and $\omega_q=\nu_{LA}\,q$ the $\Gamma$-LA phonon frequency.\\

\paragraph*{Scattering rates:} Having determined the many-particle Hamilton operator, we can evaluate the Heisenberg equation of motion $i\hbar \frac{d}{dt} \mathcal{O}(t)=[\mathcal O(t),H]$ to determine the temporal evolution of any observable $\mathcal{O}(t)$. To describe the carrier dynamics, we need equations of motion for the microscopic polarization $p_{\bm k}(t)$, carrier occupations $\rho^\lambda_{\bm k}(t)$, and phonon occupation $n^j_{\bm q}(t)$. Many-particle interactions  couple the
dynamics of these single-particle quantities to
higher-order terms describing the correlation between carriers.  
The resulting set of equations is not closed and an
infinite hierarchy of quantities with increasing number of involved particles appears \cite{haug04}. 
In this work, we apply  the correlation expansion and consider only contributions up to the second order assuming that higher-order terms are negligibly small \cite{fricke96}. This factorization technique leads to a closed set of equations for the single-particle elements. This is called second order Born approximation \cite{kira06}. 
Furthermore, we apply the Markov approximation, which neglects quantum-mechanic memory effects and accounts for a conservation of energy \cite{malic11b, malic13}.
Since we focus on the relaxation dynamics of optically excited carriers close to the Dirac point, excitonic effects  play a minor role and will be neglected. Energy renormalization stemming from Hartree-Fock contributes has already been taken into account in the calculation of the electronic dispersion relation.  

In the limit of the second-order Born-Markov approximation, we obtain the graphene Bloch equations  that have been introduced in the main part of the manuscript, cf. Eqs. (\ref{bloch_p})-(\ref{bloch_nph}). The appearing many-particle (MP) scattering processes are described by the Boltzmann-like scattering equation (Eq. (\ref{bloch_rho}))
\begin{equation}
\label{boltzman}
 \dot{\rho}_{\bm  l}(t)\big{|}_{MP}=\Gamma_{\bm  l}^{in}(t)\left(1-\rho_{\bm l}(t)\right)-\Gamma_{\bm l}^{out}(t)\rho_{\bm  l}(t)
\end{equation}
with time- and momentum-dependent scattering rates $\Gamma_{\bm 
  l}^{in/out}(t)$. The latter  include both carrier-carrier ($cc$)
and carrier-phonon ($cp$) relaxation channels, i.e. 
$\Gamma_{{\bm l}}^{in/out}(t)=\Gamma_{{\bm
    l}, cc}^{in/out}(t)+\Gamma_{{\bm l}, cp}^{in/out}(t)$. 

In the case of Coulomb-induced scattering, the in- and out-scattering rates read
\begin{equation}
 \Gamma_{{\bm l},cc}^{in/out}(t)=\frac{2\pi}{\hbar}\sum\limits_{{\bm l}_1,{\bm l}_2,{\bm l}_3}
\tilde{V}^{{\bm l},{\bm l}_1}_{{\bm l}_2,{\bm l}_3}
\mathcal{R}^{in/out}_{cc}(t)\,\delta\left(\Delta \varepsilon^{\bm l, \bm l_1}_{\bm l_2, \bm l_3}\right)
\label{ccrate}
\end{equation}
with $\tilde{V}^{{\bm l},{\bm l}_1}_{{\bm l}_2,{\bm l}_3}=V^{{\bm l},{\bm l}_1}_{{\bm l}_2,{\bm l}_3}(2V^{{\bm l},{\bm l}_1\,*}_{{\bm l}_2,{\bm l}_3}-V^{{\bm l}_1,{\bm l}\,*}_{{\bm l}_2,{\bm l}_3})$ and $\Delta \varepsilon^{\bm l, \bm l_1}_{\bm l_2, \bm l_3}=(\varepsilon_{{\bm l}}+\varepsilon_{{\bm l}_1}-\varepsilon_{{\bm l}_2}-\varepsilon_{{\bm l}_3})$.
The influence of  Pauli blocking is explicitly included in the terms $\mathcal{R}^{in}_{cc}(t)=\left(1-\rho_{{\bm
    l}_1}(t)\right)\rho_{{\bm l}_2}(t)\rho_{{\bm l}_3}(t)$ and $\mathcal{R}^{out}_{cc}(t)=\rho_{{\bm l}_1}(t)\left(1-\rho_{{\bm
    l}_2}(t)\right)\left(1-\rho_{{\bm l}_3}(t)\right).$
The
efficiency of scattering channels is determined by the screened Coulomb matrix
elements $V^{{\bm l}\;,{\bm l}_1}_{\,{\bm l}_2,{\bm l}_3}$ and the time-dependent occupation probabilities of the involved states.   
The appearing delta function  results
from the Markov approximation and allows only scattering processes,
which fulfill the conservation of energy.

The corresponding phonon-induced in-scattering rates read
\begin{eqnarray}
\label{rate_cp}
&&\Gamma_{\lambda {\bm k}, cp}^{in}(t)=\frac{2\pi}{\hbar}\sum\limits_{\lambda',\gamma,j, {\bm q}}\left|g_{{\bm q},j}^{\bm k+\bm q \lambda', \bm k \lambda}\right|^2\times
\\\nonumber
&&\rho_{{\bm k}+{\bm q}}^{\lambda'}(t)
\left[\left(n^j_{{\bm q}}(t)+1\right)\,\delta\left(\Delta \varepsilon^{\lambda\lambda',-}_{{\bm k}, \bm q, j}\right)+n_{{-\bm q}}^{j}(t)\, \delta\left(\Delta \varepsilon^{\lambda'\lambda, +}_{{\bm k}, \bm q, j}\right)\right]
\end{eqnarray}
with the condition for the conservation of energy 
$\Delta \varepsilon^{\lambda'\lambda, \pm}_{{\bm k}, \bm q, j}=(\varepsilon_{{\bm k}+{\bm q}}^{\lambda'}-\varepsilon_{{\bm k}}^{\lambda}\pm\hbar\omega_{j\bm{ q}})$ including the emission and absorption of phonons. The latter depends on the phonon occupation $n^j_{\bm q}(t)$, while the phonon emission scales with $(n^j_{\bm q}(t)+1)$ and therefore can take place at any temperature.
An excited electron scatters from the state $(\lambda', \bm k+q)$ into the state $(\lambda, \bm k)$. The momentum and energy conservation is fulfilled by emitting or absorbing a corresponding phonon.
The out-scattering rate
$\Gamma_{\lambda {\bm k}, cp}^{out}$ is obtained by substituting
$\rho_{\bm l}\leftrightarrow(1-\rho_{\bm l})$ and $n_{\bm
  u}\leftrightarrow(n_{\bm u}+1)$ in Eq. (\ref{rate_cp}).

Scattering via phonons can be very efficient and lead to the generation of hot phonons. The absorption of the latter through the electronic system can give rise to a considerable slow-down of the relaxation dynamics \cite{butscher07}. Therefore, it is very important to go beyond the bath approximation and to explicitly consider the dynamics of the phonon occupation $n^j_{\bm q}(t)$. In analogy to the carrier population in Eq. (\ref{boltzman}), we obtain the Boltzmann-like scattering equation for phonon occupations (Eq. (\ref{bloch_nph})):
\begin{equation}
 \dot{n}_{\bm u}(t)\big{|}_{MP}=\Gamma_{\bm u}^{em}(t)\big(n_{\bm u}(t)+1\big)-\Gamma_{\bm u}^{ab}(t)n_{\bm u}(t).
\end{equation}
with the phonon emission rate 
\begin{eqnarray}
\label{ph-rate}
  \Gamma_{{\bm
      q}, j}^{em}(t)&=&\frac{2\pi}{\hbar}\sum\limits_{\lambda,\lambda',{\bm
      k}}
\left|g_{{\bm q}, j}^{\bm k+\bm q \lambda', \bm k \lambda}\right|^2\times\\\nonumber
&&
\rho_{{\bm k}+{\bm q}}^{\lambda}(t)\left(1-\rho_{{\bm k}}^{\lambda'}(t)\right)\,\delta\left(\Delta \varepsilon^{\lambda'\lambda, -}_{{\bm k}, \bm q, j}\right).
\end{eqnarray}
The efficiency of phonon emission depends on the square of the carrier-phonon coupling element $g_{{\bm q}, j}^{\bm k, \lambda\lambda'}$ as well as on the occupation of the initial state $\rho_{{\bm k}+{\bm q}}^{\lambda}(t)$, and the availability of an empty final state $\rho_{{\bm k}}^{\lambda'}(t)$.
The corresponding phonon absorption rate is obtained by substituting $\rho^\lambda_{\bm
 k +\bm q}$ by $(1-\rho^\lambda_{\bm k +\bm q})$ and $(1-\rho^{\lambda'}_{\bm
 k})$ by $\rho^{\lambda'}_{\bm k}$.

Finally, many-particle interactions do not only change the occupation probability of the involved states, they also induce an ultrafast dephasing of the microscopic polarization:
\begin{equation}
 \dot{p}_{\bm k}(t)\big{|}_{MP}=-\mathcal{\gamma}_{2, \bm k}(t)p_{\bm k}(t)+\mathcal{U}_{\bm k}(t)\label{eq:pBoltz}
\end{equation}
consisting of a non-diagonal $\mathcal{U}_{\bm k}(t)$ and a diagonal part $\mathcal{\gamma}_{2, \bm k}(t)$. 
The latter is given by the time- and momentum-dependent Coulomb- and phonon-induced scattering rates  via
\begin{equation}
\label{diagonal}
 \mathcal{\gamma}_{2,\bm
  k}(t)= \frac{1}{2}\sum_{\lambda}\left(\Gamma_{\lambda,\bm
  k}^{in}(t)+\Gamma_{\lambda,\bm k}^{out}(t)\right)\,.
\end{equation}
The off-diagonal dephasing  couples to all coherences in the entire Brillouin zone yielding
\begin{equation}
\label{off-diagonal}
 \mathcal{U}_{\bm k}(t)=\sum_{\bm k'}\left(T^a_{\bm k, \bm k'}(t)p_{\bm k'}(t)+T^b_{\bm k, \bm k'}(t)p^*_{\bm k'}(t)\right)\,.
\end{equation}
The contribution stemming from the Coulomb interaction reads
\begin{eqnarray}
\nonumber
 T^i_{\bm k, \bm k'}(t)&&=\frac{\pi}{\hbar}\sum_{\bm l_1, {\bm l}_2, \lambda}\bigg(\hat{V}^{\bm k\, c,\,\,{\bm l}_2}_{{\bm k'} \lambda_i,{\bm l}_1}\,\hat{V}^{\bm k' \lambda'_i,{\bm l}_1}_{{\bm k} \,v,\,\,\,\,{\bm l}_2}\,
\mathcal{R}(t)\,
\delta(\varepsilon^{\lambda}_{\bm k}\mp\varepsilon_{\bm k'}^{\lambda}-\varepsilon_{{\bm l}_1}+\varepsilon_{{\bm l}_2})\\
&&-V^{\bm k c,{\bm k'} \lambda'_i}_{{\bm l}_2,{\bm l}_3}\,\hat{V}^{{\bm l}_2,{\bm l}_3}_{{\bm k} v,{\bm k'} \lambda_i}\,
\mathcal{\tilde{R}}(t)
\,\delta(\varepsilon^{\lambda}_{\bm k}\mp\varepsilon_{\bm k'}^{\lambda}-\varepsilon_{{\bm l}_1}-\varepsilon_{{\bm l}_2})
\bigg)
\end{eqnarray}
with $\lambda_i=c,\, \lambda'_i=v\,\,\,$  ($\lambda_i=v, \,\lambda'_i=c$) and $-$ ($+$) in the delta function in the case of $T^a_{\bm k, \bm k'}$ ($T^b_{\bm k, \bm k'}$). For reasons of clarity, we introduced the abbreviation $\hat{V}^{\bm   l_1,{\bm l}_2 }_{{\bm l}_3,{\bm l}_4}\equiv V^{\bm   l_1,{\bm l}_2 }_{{\bm l}_3,{\bm l}_4}-V^{\bm   l_2,{\bm l}_1 }_{{\bm l}_3,{\bm l}_4}$, $\mathcal{R}=(1-\rho_{{\bm
    l}_1})\rho_{{\bm l}_2}\rho_{{\bm k}}^\lambda+\rho_{{\bm
    l}_1}(1-\rho_{{\bm l}_2})(1-\rho_{{\bm k}}^\lambda),$ and $ \mathcal{\tilde{R}}=(1-\rho_{{\bm
    k}}^\lambda)\rho_{{\bm l}_1}\rho_{{\bm l}_2}+\rho_{{\bm
    k}}^\lambda(1-\rho_{{\bm l}_1})(1-\rho_{{\bm l}_2})$.
The contribution of the   carrier-phonon scattering to off-diagonal dephasing can be obtained in a similar way and reads
\begin{eqnarray}
 \mathcal{U}_{\bm k}(t)&=&\frac{\pi}{\hbar}\sum\limits_{\bm{q}\lambda j}\left|g^{{\bm k}+{\bm q}\lambda{\bm k}\lambda}_{\bm{q}j}\right|^2\times \\
&&\Bigg(
\mathcal{S}(t)
\,\delta\left(\Delta \varepsilon^{\lambda\lambda',-}_{{\bm k}, \bm q, j}\right)+
\mathcal{\tilde{S}}(t)
\,\delta\left(\Delta \varepsilon^{\lambda\lambda',+}_{{\bm k}, \bm q, j}\right)\Bigg)\nonumber
\end{eqnarray}
with 
$\mathcal{S}(t)=\left[\left(1-\rho_{\bm{k}}^{\lambda}(t)\right)\left(n_{\bm{q}}^j(t)+1\right)+\rho_{\bm{k}}^{\lambda}(t)\, n_{\bm{q}}^j(t)\right]\,p_{{\bm k}+{\bm q}}(t)$ and 
$\mathcal{\tilde{S}}(t)=\big[(1-\rho_{\bm{k}}^{\lambda})n_{-\bm{q}}^j+\rho_{\bm{k}}^{\lambda}(n_{-\bm{q}}^j+1)\big]p_{{\bm k}+{\bm q}}$.

The time- and momentum-dependent Coulomb- and phonon-induced scattering rates as well as the dephasing of the microscopic polarization have been taken fully into account for the evaluation of the graphene Bloch equations.

\bibliographystyle{andp2012}
\bibliography{papers}

\end{document}